\documentclass[preprint]{aastex6}
\usepackage{graphics,graphicx}
                                                                                     
\begin{document}                                                                     
%\baselineskip=11pt       \documentclass[preprint]{aastex6}
%\documentclass[12pt,preprint]{aastex}
%\documentstyle[12pt,/home/evans/pap/aasms4]{article}
%\documentstyle[11pt,aaspp4]{article}
%\documentstyle[aas2pp4]{article}

%\documentstyle[11pt,eqsecnum,aaspp4]{article}                                                           
%\title{{\it XMM-Newton} Observations of $\eta$ Aql \altaffiliation{1} }
\title{X-rays in Cepheids:  Identifying Low-Mass Companions of 
Intermediate-Mass Stars
\footnote{Based on observations obtained with  the Chandra X-ray Observatory} 
} 

%\author{Nancy Remage Evans,\altaffiliation{2}
%H. Moritz G\"unther,\altaffiliation{2}
%Howard E. Bond,\altaffiliation{3}
%Gail H. Schaefer,\altaffiliation{4} 
%Brian D. Mason,\altaffiliation{5} 
%Margarita Karovska,\altaffiliation{2}  
%and   
%Evan Tingle\altaffiliation{2}  

\author{Nancy Remage Evans}
\affil{Smithsonian Astrophysical Observatory,
MS 4, 60 Garden St., Cambridge, MA 02138; nevans@cfa.harvard.edu}

\author{Scott  Engle}
\affil{Department of Astronomy and Astrophysics, Villanova University, 800 Lancaster Ave., 
Villanova, PA, 19085, USA}

\author{Ignazio Pillitteri}
 \affil{INAF-Osservatorio di Palermo, Piazza del Parlamento 1,I-90134 Palermo, Italy}

\author{Edward Guinan}
\affil{Department of Astronomy and Astrophysics, Villanova University, 800 Lancaster Ave., 
Villanova, PA, 19085, USA}

\author{H. Moritz G\"unther}
\affil{Massachusetts Institute of Technology, Kavli Institute for Astrophysics and
Space Research, 77 Massachusetts Ave, NE83-569, Cambridge MA 02139, USA}

\author{Scott Wolk}
 \affil{Smithsonian Astrophysical Observatory,
MS 4, 60 Garden St., Cambridge, MA 02138}

\author{Hilding Neilson}
 \affil{Department of Astronomy and Astrophysics, University of Toronto, 
50 St. George Street, Toronto, ON, Canada M5S3H4}

\author{Massimo Marengo}
 \affil{Department of Physics and Astronomy, Iowa State University, Ames, IA, 50011,
 USA}

\author{Lynn D. Matthews}
 \affil{Massachusetts Institute of Technology, Haystack Observatory, 99 Millstone Rd., 
Westford, MA 01886, USA}

\author{Sofia Moschou}
 \affil{Smithsonian Astrophysical Observatory,
MS 4, 60 Garden St., Cambridge, MA 02138}

\author{Jeremy J. Drake}
 \affil{Smithsonian Astrophysical Observatory,
MS 4, 60 Garden St., Cambridge, MA 02138}

\author{Elaine M. Winston}
\affil{Smithsonian Astrophysical Observatory,
MS 4, 60 Garden St., Cambridge, MA 02138; nevans@cfa.harvard.edu}

\author{Maxwell Moe}
 \affil{University of Arizona, Steward Observatory, 933 N. Cherry Ave., Tucson, AZ 85721, USA}   

\author{Pierre Kervella}
 \affil{LESIA, Observatoire de Paris, Universit\'e PSL, CNRS, Sorbonne Universit\'e, Universit\'e de Paris, 5 Place Jules Janssen, 92195 Meudon, France}

\author{Louise Breuval}
 \affil{LESIA, Observatoire de Paris, Universit\'e PSL, CNRS, Sorbonne Universit\'e, Universit\'e de Paris, 5 Place Jules Janssen, 92195 Meudon, France}
 \affil{ Department of Physics and Astronomy, Johns Hopkins University, Baltimore, MD 21218, USA }

\begin{abstract} X-ray observations have been made of a sample of 20 classical Cepheids,
 including two new observations (Polaris and {\it l} Car) reported here. 
 The occurrence of X-ray flux around 
the pulsation cycle is discussed.  Three Cepheids are detected ($\delta$ Cep,
$\beta$ Dor, and Polaris).
X-rays have also been detected from the
 low--mass F, G, and K companions
of 4 Cepheids (V473 Lyr, R Cru, V659 Cen, and W Sgr), and one hot companion (S Mus).  
Upper limits on the X-ray flux of the remaining  Cepheids provide an estimate that  28\% 
have low mass companions.  This fraction of low--mass 
companions in intermediate mass Cepheids is significantly lower than expected from random
pairing with the field IMF.  Combining the companion fraction from X-rays with 
than from ultraviolet observations results in a binary/multiple fraction 
of 57\% $\pm$12\% for Cepheids with the  ratios q $>$ 0.1 and  separations a $>$ 1 au.  
This is  a lower limit since M stars are not included.
X-ray observations detect less massive companions than other 
existing studies of  intermediate mass stars.
 Our measured occurrence rate of unresolved, 
low-mass companions to Cepheids suggests that intermediate-period binaries derive from a 
combination of disk and core fragmentation and accretion. This yields a hybrid mass-ratio 
distribution that is skewed toward small values compared to a uniform distribution but 
is still top-heavy compared to random pairings drawn from the IMF.

\end{abstract}



%% In the first two sections, notice the use of the natbibcitep

\section{Introduction}

Massive and intermediate mass stars typically form as members of a pair or group.  This is an
important aspect, for instance, of the evolution of angular momentum in the pre-main sequence
phase.  
Many exotic objects in later phases of evolution arise from the combination of a compact object
in a binary or multiple system.  For example, this combination produces core collapse supernovae and 
even  gravitational wave systems.  Cepheids are most commonly approximately 5 M$_\odot$ stars, 
 intermediate mass 
stars rather than a high mass stars.  They typically ultimately become white dwarfs, 
although the most massive may become  neutron stars.  However,
their binary/multiple characteristics are similar to those of more massive stars, and can provide
insight into evolution past the main sequence.  Cepheid progenitors are B stars. 
Banyard, et al (2021) provide a recent summary of B star binary properties for comparison with 
Cepheid properties.

Components of stars in a multiple system can be challenging to disentangle.  
Intermediate mass Cepheids provide good examples of the many approaches needed to derive
the masses and separations of the components.  
Radial velocity studies
of spectroscopic binaries (Evans et al. 2015) and 
high resolution techniques provide basic information
(Evans, et al. 2020a), supplemented by 
proper motions in the Gaia era (Kervella, et al. 2019 a, b).   
However, in  multiple systems additional information is frequently 
needed to identify all the system components.  For  Cepheids, the fact that they
have evolved into cool supergiants means it is possible to identify a complete list of hot companions
in ultraviolet spectra (Evans et al. 2013) with spectral types of B and early A
(called ``late B stars'' below).  Low mass companions, 
however, are more elusive, since the spectrum at ultraviolet, optical, and infrared 
wavelengths is dominated by the more 
luminous supergiant.  X-rays provide a good remedy for this problem.  

Cepheids, like other coronal  supergiants  (Engle 2015; Ayres 2011) produce a comparatively modest X-ray
flux.  $\delta$ Cep itself typically has an X-ray luminosity log L$_X$ = 28.6 ergs sec$^{-1}$.  
However, in an exciting development Engle et al. (2017) found a sharp increase in X-ray flux 
for a brief period near maximum radius in the pulsation cycle.  This was seen in two pulsation 
cycles and also in the Cepheid $\beta$ Dor.

Main sequence stars of spectral types F, G, and K at the age of Cepheids (typically 50 Myr), 
on the other hand are much more vigorous 
producers of X-rays.  This makes X-rays a 
good discriminant  between young physical companions of Cepheids and old field stars. 
Mapping  the X-ray production of low--mass main sequence stars in  
 temperature and age has been an important contribution of X-ray studies.  We use this legacy to
predict X-ray fluxes from possible companions at the age of Cepheids.  Details are discussed 
in Section 4.4.

The vigorous X-ray production of low-mass main sequence stars
adds an important piece to the determination of the properties of the multiple systems 
of Cepheids and other intermediate mass stars.  
X-ray observations of Cepheids where the upper limit is below the level of possible
main sequence companions indicate that a low-mass companion is {\it not} present.
This provides the fraction of Cepheid systems with low-mass companions. 
 Since low-mass stars dominate the stellar 
mass distribution, identifying them is  important  for putting together a 
 complete picture of star formation.  

A thorough discussion of the observed properties of binary and multiple systems
systems is given in Moe and 
Di Stefano (2017).  In particular they discuss the distributions of mass ratios and separations 
as a function of mass of the primary and the implications for star formation.  
The distribution of mass ratios as a function of 
separation for O and B stars divides into 3 separation regimes.  Close binaries with separations
$<$ 0.4 au favor reasonably massive companions, with q = M$_2$/ M$_1$ $\simeq$ 0.5.    
In this separation range presumably competitive accretion has resulted in relatively equal masses 
of the components.   Stars
in this separation range are not present in the Cepheid sample due to Roche lobe overflow (RLOF).
Systems with wider separations up to 200 au tend to have smaller mass ratios 
 q $\simeq$ 0.2 to 0.3. Companions at wider separations (200 to 5000 au) in O B systems tend
to be outer components in triple systems.  Their masses are close to a random pairing with 
the IMF, favoring low-mass companions.  Fig. 1 in Moe and Di Stefano shows
that the addition of systems with mass ratios q as small as 0.1 in the 
present study fills a gap for  stars
as massive as O and B stars.   X-rays are the one spectral region where low-mass main sequence
companions can be detected, since at other wavelengths the supergiants outshine dwarfs.  

A useful comparison to the fraction of Cepheid plus low-mass systems 
is the determination of low--mass companions of B and early A stars since these 
are the stars that evolve into Cepheids.  A Chandra observation of the cluster Trumpler 16 was 
used to identify X-ray sources among  these stars (Evans, et al. 2011).  
Since B and early A stars do not typically
produce X-rays, it was assumed that low-mass companions were the X-ray producers.  They concluded 
that  39\% of these late B stars have a low--mass companion.   Two small points are of note in 
this comparison.  
Cepheids are slightly older than 
Tr 16 B stars which are $\simeq$3Myr. 
Also  this fraction in Tr 16  includes companions at all 
separations, where Cepheid binaries with separations smaller than 1 au have been removed due to
RLOF during the red giant phase.   However the Tr 16 results are a good comparison
to the Cepheid results in this study.

A second aspect of the present study is that  
   Cepheid upper atmospheres have several properties including X-rays which may bear on outstanding 
questions. Cepheids frequently have excess infrared (IR) emission from circumstellar 
envelopes (CSEs) summarized by Gallenne, et al (2021), and Hocde, et al. (2020a,b, 2021). 
This effect needs to be quantified to allow precise distance determinations
 using  the Cepheid Leavitt (Period-Luminosity) law in the IR.
In addition, the CSEs are related to the long-standing question of possible mass loss in Cepheids. 
X-ray flux controlled by 
the pulsation cycle may be a driver of  CSEs, and hence a clue to understanding
both mass loss and the IR Leavitt Law.

This study begins with new Chandra observations of two 
 Cepheids  ({\it l} Car and Polaris)  near maximum radius to determine whether they show
the flux increase observed in $\delta$ Cep.  These new data have then been combined with
archival data to investigate the occurrence of young low--mass  X-ray active companions.  

{\bf l Car} is an important Cepheid because it is bright and also has a long
pulsation period (35$^d$). Long 
period Cepheids are vital for determining distances to external galaxies.  Like many long period 
Cepheids, it has modest variation in some of its parameters such as its period.  This was 
explored in detail with a combination of radial velocities and interferometry by Anderson, 
et al. (2016).

{\bf Polaris} is the nearest and brightest Cepheid.  An ongoing program is measuring its 
mass from its astrometric orbit (Evans, et al. 2018).  
The distance to Polaris has been controversial recently.  However, the Gaia EDR3 parallax
to the resolved companion Polaris B now seems to provide a reliable value of 137 pc 
(Evans et al. 2018).
Polaris has been observed four times in X-rays,
 once by Chandra (Evans, et al. 2010) and three
times by XMM-Newton (Engle 2015). All four
observations have a reasonably constant X-ray
luminosity $\simeq$ log L$_X$ = 28.9 ergs s$^{-1}$ . 
However, none of the observations have
been made at the “phase of interest”  (maximum radius)
for comparison with the X-ray burst of $\delta$ Cep. 
To add to the information about the upper atmosphere,
HST COS spectra 
provide chromospheric emission lines which have
been analyzed by Engle (2015).
Polaris also has a radius and CSE measured by
interferometry (Merand, et al. 2006).

In this paper, subsequent sections discuss observations of {\it l} Car and Polaris at maximum radius to 
search for increased X-ray flux at this phase, and the relation of X-ray observations to the
pulsation cycle.
 The sample of X-ray observations of Cepheids is assembled from 
the new observations, upper limits from the survey investigating resolved companions,  
and observations where the Cepheids and low-mass companions were detected.  This combined sample
is compared with X-rays from main sequence stars.  Finally the discussion includes the fraction of
Cepheids in binary and multiple systems 
and the implications for star formation of  these
 intermediate mass stars with small mass ratios.

\section{Observation and Data Analysis}

In order to investigate further the X-ray flux from Cepheids,  
observations of two stars ({\it l} Car and and Polaris)
were made with Chandra. The observations were timed 
exposures with the ACIS-I instrument, and are listed in Table~\ref{obs}.  
Reductions were done with the standard {\it CIAO} software package.  
\footnote{https://cxc.cfa.harvard.edu/ciao/} 

{\bf {\it l} Car}

  The long period Cepheid {\it l} Car is known to have period fluctuations like 
other long period Cepheids (Anderson, et al, 2016a; Anderson 2016b ).  
The phases of the 
observations were computed from the period summary in Neilson, et al. (2016), including the changing 
period.  
%The phases of the 
%observations in Table 1 are computed using P = 	35.568354$^d$ and the epoch of maximum light 
%2,457,517.29 JD.  
The time of observation was selected based on the relation between maximum 
radius and the burst of X-rays in $\delta$ Cep, where  the X-ray burst occurs approximately 0.1 in 
phase after maximum radius.  The phase of maximum radius has been measured in 3 successive
cycles by Anderson, et al. (2016a), providing a recent determination of the time of maximum radius.  
The phase of  R$_{max}$ is 0.40, leading to a requested phase of the X-ray observation of 0.50.  
The phases of observations are listed in Table~\ref{obs}.  The observation had to be 
broken into two parts for scheduling reasons, as listed in Table~\ref{obs}.  Note that for a period of
35$^d$, the duration of the longer exposure covers only 0.02 in phase,  

No source was detected at the position of the Cepheid {\it l} Car.  The upper limit to the flux was 
estimated as follows, using 
E(B-V) = 0.17 mag (Fernie, et al. 1995), the conversion to N$_H$ from Seward 
(2000; N$_H$/E(B-V) = 5.9$\times$10$^{21}$ atoms cm$^{-2}$ mag$^{-1}$)
 and a distance of 506 pc (Evans et al 2016a). 
Distances for Cepheids (except Polaris) are taken from Evans, et al., (2016a) based on the 
HST FGS scale of Benedict, et al (2007). 
Less than 1 count was found for {\it l} Car
in 58.3 ksec. Using PIMMS, this provides an unabsorbed flux  of 
5.88 x 10$^{-16}$ ergs s$^{-1}$ cm$^{-2}$.  At the distance of the Cepheid, this corresponds 
to a luminosity L$_X$ of 1.75 x 10$^{28}$ ergs s$^{-1}$ (log L$_x$ = 28.26). The luminosity upper 
limit for the short exposure is  log L$_x$ = 28.70  and for the combined exposure    
log L$_x$ = 28.11.  Uncertainties on the upper limits are based on the variance of 
the background.  For {\it l} Car, the  exposure time  (corrected to 44 ksec of good time 
intervals) corresponds to a 9\% uncertainty, 
or a difference of 0.04 in log L$_X$.  

{\bf Polaris}

  For Polaris, similarly, the observation was requested to coincide with the predicted time of 
X-ray increase shortly after maximum radius.   
Phases of observation were computed from the recent ephemeris (Engle 2015):
2,455,909.910 + 3.972433 E.  
The phase range covered by the observation is  0.48 to 0.69.

A source was detected at the position of Polaris.  A flux was determined by fitting the
spectrum with a MEKAL model in {\it CIAO} with E(B-V) = 0.00 and a fixed temperature 
of 0.56 keV, typical of a young star.  MEKAL models were used for consistency with 
data from Engle (2015), however tests with newer APEC models agreed to  10\%.     
The resulting flux is 2.08 x 10$^{-14}$ ergs s$^{-1}$ cm$^{-2}$.
The distance used to determine the luminosity is
137 pc from the Gaia EDR3 distance to the resolved companion Polaris B.  Because
of the length of the exposure, the luminosity was computed for two halves,  
%%%%%  The corresponding luminosity L$_X$ is 4.67 x 10$^{-28}$ ergs s$^{-1}$
 log L$_X$ 28.83 and 28.71 [ergs s$^{-1}$].  The details are listed in Table~\ref{obs}.

%xxxx ROSAT points on figure ???  put in table, upper limit figure

%l car 
% 20149 &	ACIS-I &   58.27 2018-09-27 03:54:25   2458388.6627   58388.9999  .51

% 21858 	ACIS-I   20.8 	2019-01-11 05:21:36   2458494.7233   58494.8437   .48

%polaris 

%18928    ACIS-I   69.16   2017-07-07 05:57:09  2457941.7480   57942.1482

% phases usig neilson et al

\begin{deluxetable}{llccccc}
%\footnotesize
\tablecaption{ Chandra Observations \label{obs}}
\tablewidth{0pt}
\tablehead{
 \colhead{OBSID} &  \colhead{Instrument}  & \colhead{Exp}  &  \colhead{JD}  & \colhead{Phase}  
  & \colhead{D}    & \colhead{log L$_X$} \\
 \colhead{} &  \colhead{}  & \colhead{ksec}  &  \colhead{mid}  & \colhead{}   & \colhead{pc}
& \colhead{  [ergs s$^{-1}$]  } \\
}
\startdata
{\it l} Car & & & & & & \\
 20149 &	ACIS-I &   58.27   & 2,458,388.9999 & 0.53 & 506  & $<$28.26 \\
 21858 &	ACIS-I &  20.8     &  2,458,494.8437 &  0.51  &  & $<$28.70  \\
Polaris  & & & & & & \\
18928   & ACIS-I &  69.16 &   2,457,942.1482  & 0.48-0.59 & 137  & 28.83   \\
        &        &        &                  &  0.59-0.69  &      &  28.71   \\
\enddata

\end{deluxetable}

\begin{figure}
\plotone{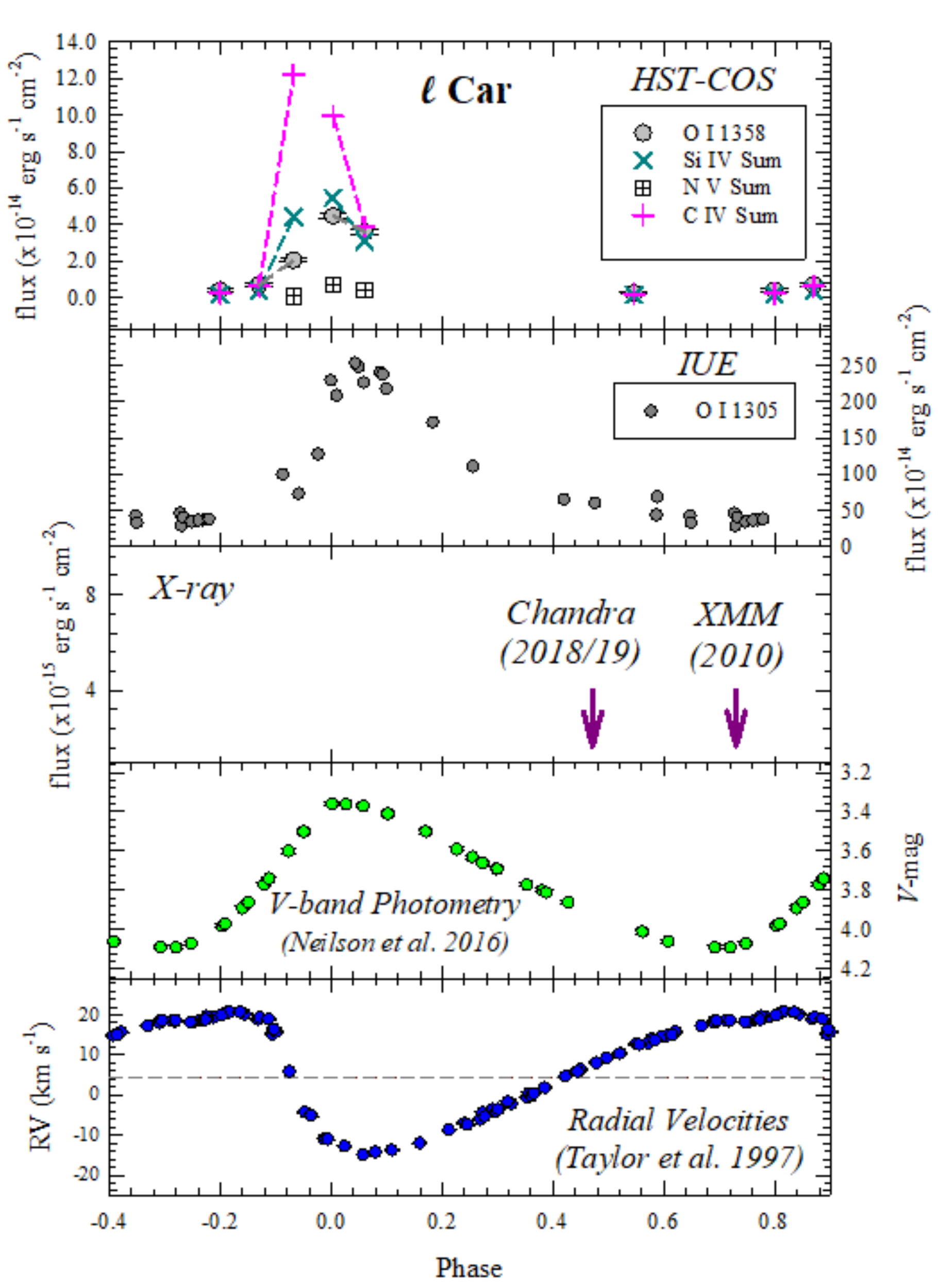}
\caption{Multiwavelength observations of {\it l} Car as a function of pulsation phase. 
Panels top to bottom show emission lines from HST COS spectra, emission lines from 
IUE spectra, X-ray observations, V photometry, and radial velocities as discussed in
the text.    
\label{lcar}}
\end{figure}

\begin{figure}
%\plotone{Polaris_X-ray_Variability_New.pdf}
% \includegraphics[width=5.0in]{polaris_xray_variability_eng.pdf}{\centering}
 \includegraphics[width=5.0in]{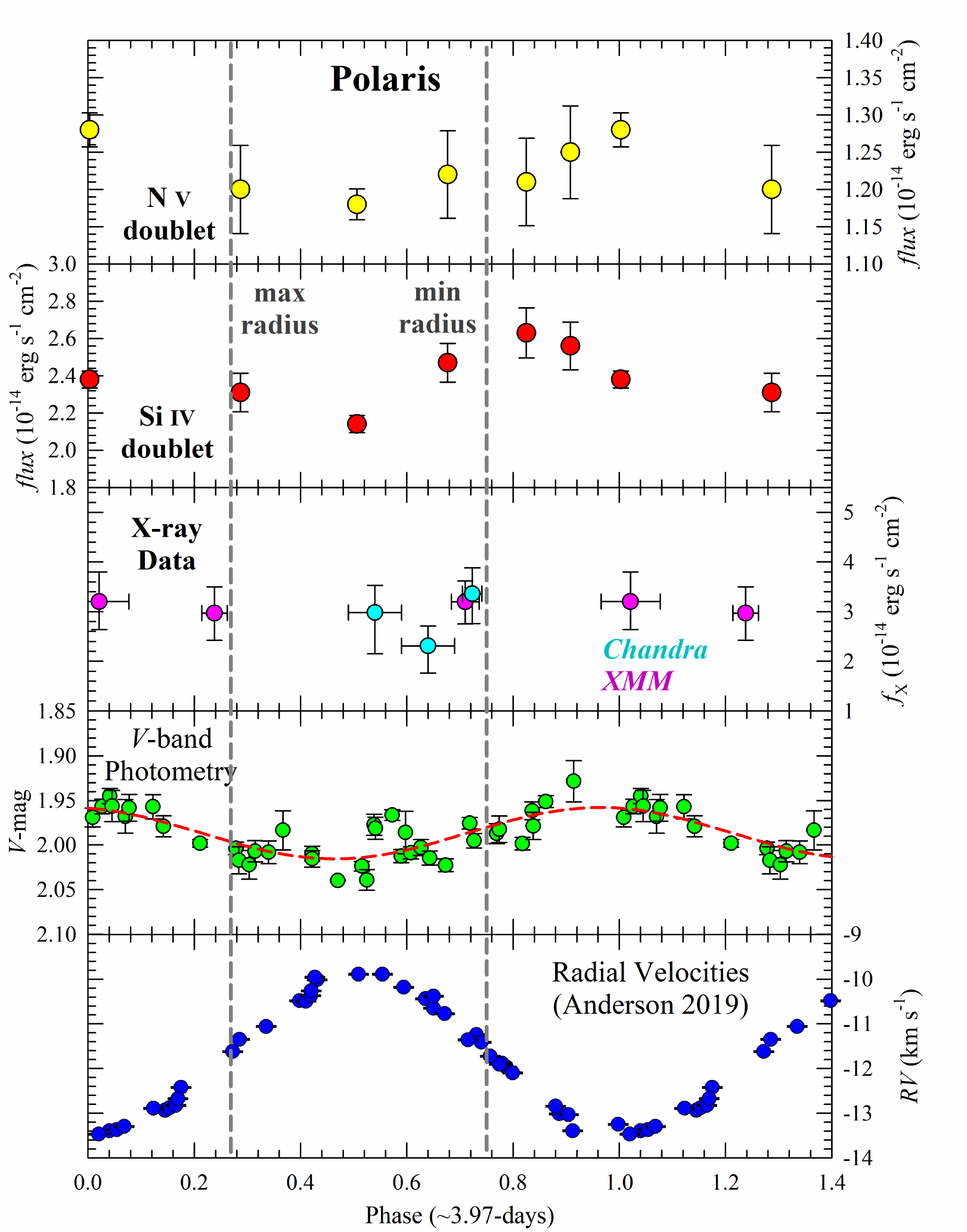}{\centering}
\caption{Multiwavelength observations of Polaris as a function of pulsation phase. 
Panels top to bottom show N V emission lines and Si IV emission lines 
from HST COS spectra,
 X-ray observations, V photometry, and radial velocities as discussed in
the text. The error bars on the X-ray phases indicate the time period covered
by the observations.  
 The Chandra observation in Table ~\ref{obs} has been broken into 2 parts. 
\label{pol}}
\end{figure}

%ephemeris--thesis

%polaris 2455909.910 + 3.972433(520) x E
%l car 2437751.5 + 35.535 E         Neilson      paper  2016

%*** using E = 57517.29  jd 
%Period: 	35.568354 	 
%xxxcheck in office for source of pd

% old: P pul  35.567748 d  T max  2,457,162.61 JD   l car prop

%Oct 15 2021: Neilson, et al changing period: 
%l car epoch 580
%P(580) = 35.554938
%e(580) = 2458370.1096

%Berdnikov etal 2003 JAAVSO,31, 146  : 2440736.9  +  35.539 E

%Szabados  1989  2440736.230 + 35.551341

%Feinstein, A and Muzzio, J 1969, A\&A, 3, 388  35.5330 +/- .00084
%    To = 2436436.0
%   used by Neilson et al

%Breitfelder et al 2016  47774.310 +  35.551609 +/- 0.000265   E
%    C=2450405.8306 +/-0.0344 + (35.556234 +/- .000373) x  E + (3.030 x 10−5 +/-.104 x 10−5) x E
%   Fig 4  omc 2450405.8306 +  35.556234 E

%Engle thesis 

%del cep   xmm

%bet dor   xmm

%Polaris   xmm, chan

%l car  

%SU Cas

\section{The Pulsation Cycle}

The interest in X-ray production in Cepheids is enhanced by its relation to other parameters 
of the pulsation cycle.  In the case of $\delta$ Cep (Engle, et al. 2017), a brief X-ray 
 burst was seen shortly after {\it maximum} radius.  This is in contrast to 
ultraviolet chromospheric lines which go into emission after {\it minimum} radius.  

{\bf {\it l} Car}:
The relation of the X-ray observations in Table~\ref{obs} to other pulsation parameters
is shown in Fig.~\ref{lcar}, adapted from Neilson, et al. (2016).  Successive panels show 
emission lines from   HST COS, spectra, and from
the International Ultraviolet Explorer satellite (IUE) spectra, upper limits from Chandra
 (Table~\ref{obs})  and 
XMM-Newton observations, V  band photometry (Neilson, et al., 2016) and radial velocities
(Taylor, et al 1997).

%discuss phases:  35 d cycle

{\bf Polaris}: 
The relation of the X-ray observations of Polaris to the pulsation cycle variables is shown in 
Fig.~\ref{pol}.  Successive panels show N V and Si IV emission lines from HST COS spectra (Engle 2015),
X-ray observations, V band photometry (Engle 2015 plus some additional data from the same system), 
and radial velocities (Anderson 2019). 
Four previous
observations with XMM-Newton and Chandra are listed in Engle (2015).

%Beta dor:  Engle thesis but then separate Engle paper with new data

\section{X-ray Observations of Cepheids}

In addition to the observations of {\it l} Car and Polaris, X-ray observations have been made of a 
number of Cepheids.   $\delta$ Cep, $\beta$ Dor, 
   {\it l} Car, and SU Cas are discussed in Engle (2015).
Observations of V473 Lyr and $\eta$ Aql are discussed in 
Evans, et al. (2020b) 
and Evans, et al. (2021) respectively. Finally, a survey was made with XMM-Newton of possible
resolved companions (Evans et al. 2016b). In this section we discuss these observations in 
three subsections: upper limits in cases where the Cepheid was not detected
 (Table~\ref{x.o.obs}), cases where the 
Cepheid was detected  (Table~\ref{x.c.obs}), 
and cases where a low-mass companion was detected  (Table~\ref{x.com.obs}).

\subsection{Upper Limits for Cepheids and Close Companions}

A series of observations was conducted with XMM-Newton of   Cepheids with 
possible resolved low mass companions 
(Evans, et al. 2016b).  The resolved companion candidates are separated from the 
Cepheids typically by 10$\arcsec$,
hence these observations also provide an X-ray observation of the Cepheid
and any possible close companion as well. For the resolved companions, 
because low mass stars at the age of Cepheids are X-ray active, they can be 
distinguished from old line-of-sight field stars in X-ray observations. 
 The companion candidates in the XMM-Newton survey were identified
in an HST Wide Field Camera 3 (WFC3) snapshot survey.
%are separated from the 
%Cepheids typically by 10$\arcsec$.
%hence these observations also provide an X-ray observation of the Cepheid
% as well.   
The exposure time was set 
to detect a late spectral type companion (see Section 4.4).  
Because young low mass main sequence stars are more X-ray active
than supergiant Cepheids,  in general  the observations of the 
 Cepheids themselves  are upper limits.  They are summarized
in Table~\ref{x.o.obs}. Columns in Table~\ref{x.o.obs} list the star, the satellite used, the epoch 
and period used to compute the phase where they are not provided in another source, the reference for
the period, the JD of mid exposure where it is not listed elsewhere, the pulsation phase of 
the observation, the distance D from Evans, et al (2016a) and the log of the X-ray luminosity of the 
upper limit.  The distances are on the Benedict, et al. (2007) scale. 
Where the observations covered a significant range of phases, the range is shown.  
For most observations, 
the exposure time only covered  $\pm$ 0.03 in phase.   The upper limits are based on a 3 $\sigma$ 
detection since the positions of the sources are known, as discussed in Evans, et al (2016b). 
Uncertainties for the upper limits were estimated from the standard deviation of the local
background.  As a typical example, V440 Per has an exposure time of 21 ksec and an 
uncertainty on the upper limit of 9\%, corresponding to 0.04 in log L$_X$.     

The upper limits to Cepheid X-rays are plotted in Fig.~\ref{ul}.  

In addition, upper limits to two Cepheids (SU Cas and {\it l} Car) were reported by Engle (2015)
from XMM-Newton observations which are listed in Table~\ref{x.o.obs}. 
Neither was detected.  For the short period 
 Cepheid SU Cas (P = 1.95$^d$;  the upper limit  
log  L$_X$ =  29.46  ergs s$^{-1}$ was estimated using exposure times and background rates, and 
a distance  D = 376 pc. 
For {\it l } Car the upper limit was estimated using  a distance 
506 pc to be  log  L$_X$ =  29.62  ergs s$^{-1}$. 
% Upper  X-ray  luminosity  limits  of logLX le  29.6  and  29.5  ergs/s  were estimated  for lCar  and SU  Cas,  respectively,  based  on  exposure  times,  background  count  rates

$\eta$ Aql was also observed  by XMM-Newton (Evans, et al. 2021) but not detected 
(Table ~\ref{x.o.obs}), providing an upper limit.

%star     To         P       ref   midJD      phi    d     log Lx

\begin{deluxetable}{lllcccccc}
%\footnotesize
\tablecaption{ X-ray Upper Limits of Cepheids  \label{x.o.obs}}
\tablewidth{0pt}
\tablehead{
 \colhead{}  &  \colhead{Sat}  &  \colhead{T$_0$}  & \colhead{P} & \colhead{Ref} &  \colhead{JD mid}  
& \colhead{Phase}  & \colhead{D} & \colhead{log L$_X$}   \\
 \colhead{} &  \colhead{}  &  \colhead{-2,400,000}  & \colhead{$^d$} & \colhead{} &  \colhead{-2,400,000}  
& \colhead{}  & \colhead{pc} & \colhead{ergs s$^{-1}$}   \\
}
\startdata
 & & & & & & & & \\
$\eta$ Aql  & XMM & 55856.689  & 7.177025  &  1 & 58616.02 & 0.41-0.53  & 273 &  $<$ 29.23    \\
l Car   & Chan & 37751.5 & 35.535 &     2  &   58389.00   & 0.55  &  506 &  $<$ 28.26 \\
l Car   & Chan & &  &     2  &  58494.84     & 0.52  &  506 &  $<$ 28.70 \\
l Car  & XMM & &   &    3    & 55232.31  &  0.76 &  506 &    $<$ 29.62 \\ % d 498  lx < 29.6
SU Cas & XMM & 55199.614 & 1.949330  &  3 & 55236.27 & 0.64-0.98  & 376  & $<$  29.46  \\  %d 395 lx < 29.5
V737 Cen &  XMM & 55118.3272 & 7.0659 &  4 &  56684.28    & 0.62 &  848   &  $<$ 29.40 \\
S Cru  & XMM  &   34973.495  & 4.689970 &  5 &  56525.48  &  0.34 &  724   &  $<$ 29.60 \\
X Cyg & XMM  & 43830.251 & 16.385692 &  6 & 56408.79  &  0.65 &  981 &   $<$  29.89 \\
R Mus  & XMM  & 26496.033 & 7.510159 &  5  & 56338.66  &  0.63 &   844 &  $<$ 29.48 \\
S Nor & XMM  &  44018.884 & 9.754244 &  5 &  57095.07   &  0.56 &  910 &  $<$ 29.44 \\
Y Oph  & XMM  & 39853.173 & 17.126908 &  5 & 56182.84  &  0.45 &  510 &  $<$ 29.29 \\
V440 Per & XMM  & 44551.137 & 7.572498 &  6 &  56538.55   &  0.02 & 791 &  $<$ 29.17 \\
U Sgr &  XMM  &  30117.955 & 6.745229 &  5 & 54020.65  &  0.64 &  617 &  $<$ 29.05 \\
Y Sgr  & XMM  & 40762.329 & 5.773380 &  5 &  56564.79  &  0.12 &  505 &  $<$  29.25 \\
 & & & & & & & & \\
\enddata

\end{deluxetable}

Period source: 1. Evans, et al. 2021;
2. Table~\ref{obs};
3. Engle, S. 2015;
4. Usenko, I, et al. 2013;
5. Szabados, L. 1989;
6. Szabados, L. 1991;

%references correct

%nb  periods/phases in Table 2 do not take account of changing period--but shift of .1d in 5d pd only .02

%xmm phase Tab 2 checked

\subsection{Detections of Cepheids}

\begin{figure}
\plotone{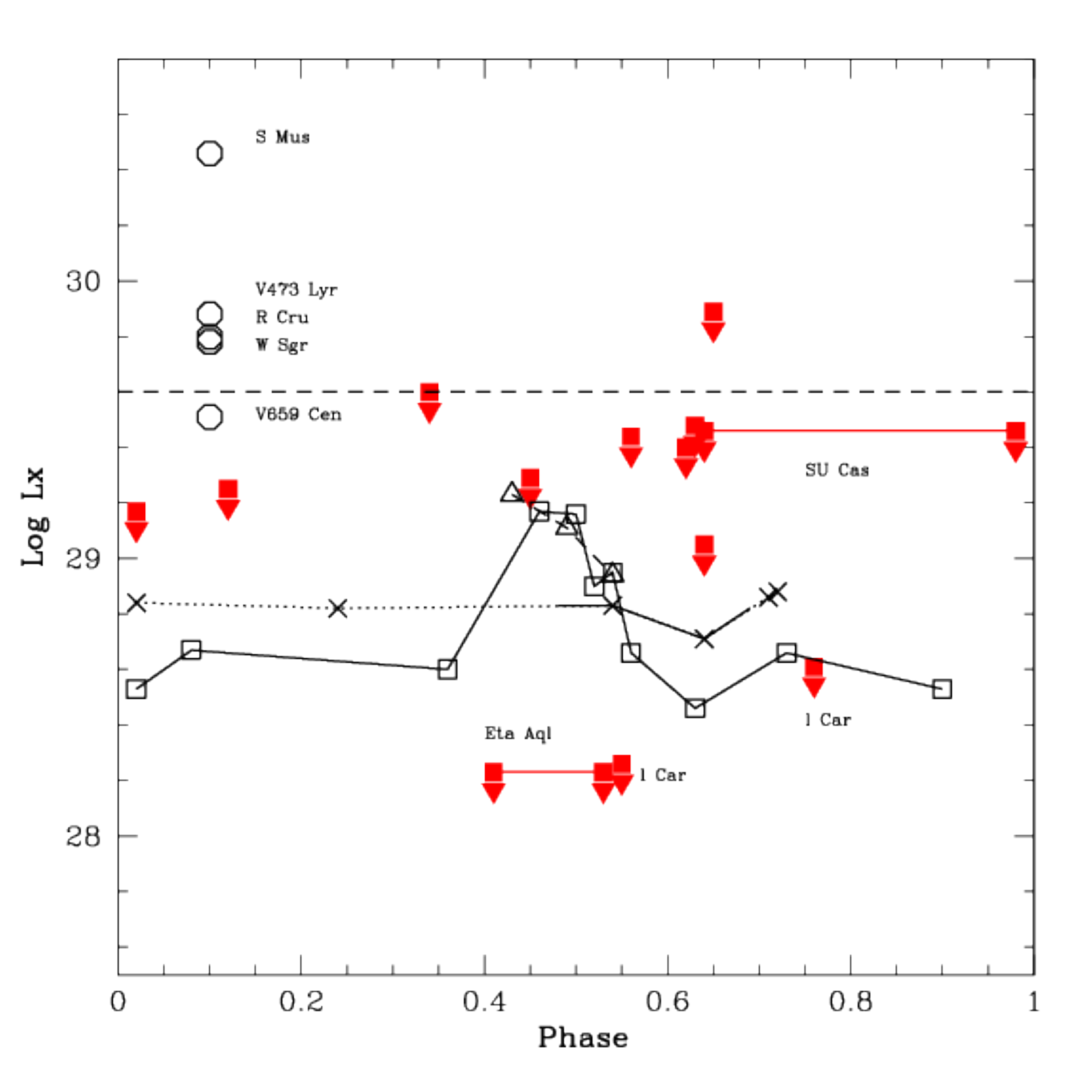}
\caption{X-ray observations of Cepheids. 
 Upper limits from non-detections in 
Table ~\ref{x.o.obs} 
are filled red down arrows; detections of Cepheids in Table ~\ref{x.c.obs}
 are: connected open squares: 
 $\delta$ Cep, connected x's: Polaris; connected triangles:  $\beta$ Dor.
The solid portion of the Polaris data line shows the phase range of the new  
observation in Table 1.
  For $\eta$ Aql and SU Cas upper limits are indicated and 
the lines show the phase range covered. 
The mean log L$_X$ for F, G, and K main 
sequence stars is shown by the dashed  line at  log L$_X$ = 29.6.
Circles in the upper left (labeled) are systems where the low-mass companion dominates the 
X-rays  (Table~\ref{x.com.obs}).   
Luminosity is in ergs s$^{-1}$. 
\label{ul}}
\end{figure}

For the Cepheids $\delta$ Cep, $\beta$ Dor, and Polaris the Cepheid itself was detected (Engle 2015).  
The observations are listed in Table~\ref{x.c.obs}  and are shown in Fig.~\ref{ul}.

There is some evidence for a low mass companion to $\delta$ Cep from radial velocities (Anderson et al. 2015)
interferometry (Gallenne, et al. 2016) and Gaia (Kervella, et al. 2019b).  While this is possible, 
Fig.~\ref{ul} shows that the X-ray level is lower than any main sequence star hotter than spectral 
type M.  

 In the 10$^d$ Cepheid $\beta$ Dor a variation in luminosity is seen, with the
largest value at about the same level as the maximum luminosity of 
$\delta$ Cep.  
 The 10$^d$ Cepheids
fall in the Hertzsprung progression of light curves where the pulsation 
amplitude is decreased by the coincidence of the primary and 
secondary humps.  This may distort the phase of maximum 
light which is the standard ephemeris fiducial.  We have determined the phase 
of the X-ray increase in $\beta$ Dor as follows.  For $\delta$ Cep (Fig. 1 in  
Engle, et al. 2017)
 both the phases when the pulsation wave passes 
through the photosphere at minimum radius and the phase of X-ray 
maximum shortly after maximum radius are well determined 
from FUV lines and X-ray  fluxes respectively. The X-ray flux
maximum occurs 0.66 phase after the FUV flux maximum. 
Similarly, the phase of FUV maximum (minimum radius)
 of the photospheric 
pulsation wave is well determined for 
$\beta$ Dor.   If we match the phases of UV maximum in $\delta$
Cep and $\beta$ Dor (by adding 0.17 to the phase of UV maximum in 
$\beta$ Dor), the phase of X-ray maximum becomes 0.42, as shown 
in Fig.~\ref{ul}
very similar to the phase of $\delta$ Cep.  This phase adjustment is 
included in the  $\beta$ Dor phases in Table~\ref{x.c.obs}.

%Del Cep:  broken into phase ranges like E17  (adjusted for Benedict distance)
%  d;  Engle: 273    Ben: 255 :  (255/273)\^2 = .872

%Beta dor: Engle 2015 thesis  pd
%d= 318   cep snap 335    (335/318)\^2  = 1.110

%Polaris:  d= 137   Engle 133
%ephemeris:  as above: recent from thesis (same as Thesis Table 18)

%xxxx   Scott E:  I cannot account for the Polaris point at about .38 phi

%xxxx:  see engle msg

%XXXXX  ALL ENTRIES IN tABLES 2 AND 3 CHECKED  XXXX

%XXXX Polaris fluxes updated from Engle Jan 11 2022  

\begin{deluxetable}{llcccc}
%\footnotesize
\tablecaption{ X-ray Detections of Cepheids  \label{x.c.obs}}
\tablewidth{0pt}
\tablehead{
 \colhead{}  &  \colhead{Sat}   & \colhead{Ref}   
& \colhead{Phase}  & \colhead{D} & \colhead{log L$_X$}   \\
 \colhead{} &  \colhead{}   & \colhead{}  
& \colhead{}  & \colhead{pc} & \colhead{ergs s$^{-!}$}   \\
}
\startdata
%detect Ceph
Polaris & XMM    & 1    & 0.21-0.26 & 137   &  28.82  \\ % 28.93    \\ % 28.90  133  (137/133)^2 = 1.061
Polaris & XMM   & 1     & 0.68-0.74 & 137  & 28.86    \\  % 28.87     \\ %  28.84
Polaris & XMM   & 1    & 0.97-0.08 & 137 & 28.90 \\ %  28.84    \\ % 28.81
Polaris & Chan   & 1    & 0.71-0.73 & 137 & 28.88 \\ % 28.93     \\ %  28.90
Polaris & Chan   &  2   & 0.48-0.59 & 137  & 28.83 \\  % 28.67    \\ %
        &        &  2   & 0.59-0.69  & 137 & 28.71  \\
$\delta$ Cep & XMM  &  3   & 0.33–0.39   & 255 &  28.60 \\   %  28.66  for 27  28.725     .872
     & XMM  &  3 &     0.43–0.48   & 255 &  29.17   \\   % 29.23
     & XMM  &  3 &     0.48–0.54    & 255 &  28.90   \\   % 28.96
      & XMM  &  3 &    0.54–0.59  & 255 &     28.66   \\ %   28.72
      & XMM  &  3 &   0.05–0.12   & 255 &   28.67   \\  % 28.73
    & XMM  &  3 &    0.84–0.96   & 255 &   28.53   \\ % 28.59
     & XMM  &  3 &   0.96–0.08   & 255 &   28.53  \\ %   28.59
     & XMM  &  3 &    0.58–0.68   & 255 &  28.46   \\ %  28.52
     & XMM  &  3 &   0.68–0.78   & 255 &  28.66  \\ %   28.72
     & Chan   &  3 &   0.48–0.52   & 255 &  29.16   \\  % 29.22
     & Chan  &  3 &   0.52–0.56   & 255 &  28.95  \\ %  29.0
%delcep &  & &  &   e17 & & .455  & &  29.23 \\
%delcep  &   & &  &   pap &    &  .51 &   &  28.96 \\
%delcep  &   & &  &   pap  &   &  .50 &  &  29.22 \\
%delcep & chan   & &  &   e17   &   & 0.48–0.52  & 273 E 255 nre &   29.22 \\
%delcep & chan   0.52–0.56                         273    255 &   29.01 \\
$\beta$ Dor & XMM   &  1 &  0.58-0.62*  & 335  & 29.24  \\ %    29.19 at d 335  1.110
$\beta$ Dor  & XMM   & 1 &  0.64-0.68* &  335 &  29.11 \\  %5   29.06 
$\beta$ Dor  & XMM    & 1 &  0.69-0.73* & 335 & 28.94  \\   % 28.89
 & & & & &  \\
\enddata

\end{deluxetable}

Sources: 1. Engle 2015; 2.  Table~\ref{obs}; 3. Engle, et al. 2017

* Phases adjusted; see text

\subsection{ Detections of Cepheid Companions}

\begin{deluxetable}{lllcccccc}
%\footnotesize
\tablecaption{ X-ray Detections Cepheid Companions \label{x.com.obs}}
\tablewidth{0pt}
\tablehead{
 \colhead{}  &  \colhead{Sat}  &  \colhead{T$_0$}  & \colhead{P} & \colhead{Ref} &  \colhead{JD mid}  
& \colhead{phase}  & \colhead{D} & \colhead{log L$_X$}   \\
 \colhead{} &  \colhead{}  &  \colhead{-2,400,000}  & \colhead{$^d$} & \colhead{P} &  \colhead{-2,400,000}  
& \colhead{}  & \colhead{pc} & \colhead{ergs s$^{-!}$}   \\
}
\startdata
%detect comp
V473Lyr & XMM  & ---* & 1.490813  & 1 & 56557.96 &  0.47 &   553 & 29.88  \\
   & XMM  & &  &  1 & 58560.00 & 0.42-0.73 &     & 30.07  \\
S Mus & XMM & 40299.163 & 9.659875 &  2  & 56298.26 & 0.24  &  789 &  30.46 \\
W Sgr  & XMM & 43374.622  & 7.594904  & 2  & 57637.64 & 0.97  &  409 & 29.78 \\
  & XMM & & & &   57659.39  &  .836  &  &  \\
V659 Cen & XMM & 52358.9089 & 5.62316689 & 3  & 56543.46 &  0.14     &    753 &  29.51 \\   
R Cru   & XMM &  55172.5100 &  5.825701 & 4  & 56662.46 &  0.73   &    829 &   29.80 \\
\enddata

\end{deluxetable}

Sources: 1. Evans et al. 2020b; 2.  Szabados 1989; 3. Berdnikov et al. 2000; 4.  Usenko et al. 2014 

* The phase for V473 Lyr very variable.

%mid JD checked jul 16 2021

In some cases X-ray observations identified low mass companions of Cepheids.  Each of those will 
be discussed in this section. They are summarized in Table~\ref{x.com.obs} and 
Fig.~\ref{ul}.

{\bf V473 Lyr}: This is a unique Cepheid with a variable amplitude, perhaps similar to the 
Blazhko effect in RR Lyrae stars.  A recent XMM-Newton observation (Evans, et al. 2020b) was 
made to follow up a possible X-ray burst.  However, the X-ray flux remained constant for a 
third of the pulsation cycle, making  a low mass companion 
the most likely interpretation. 
Limits from radial velocities and Gaia proper motions are consistent with a companion at a 
separation between 30 and 300 AU.

\begin{figure}
\plottwo{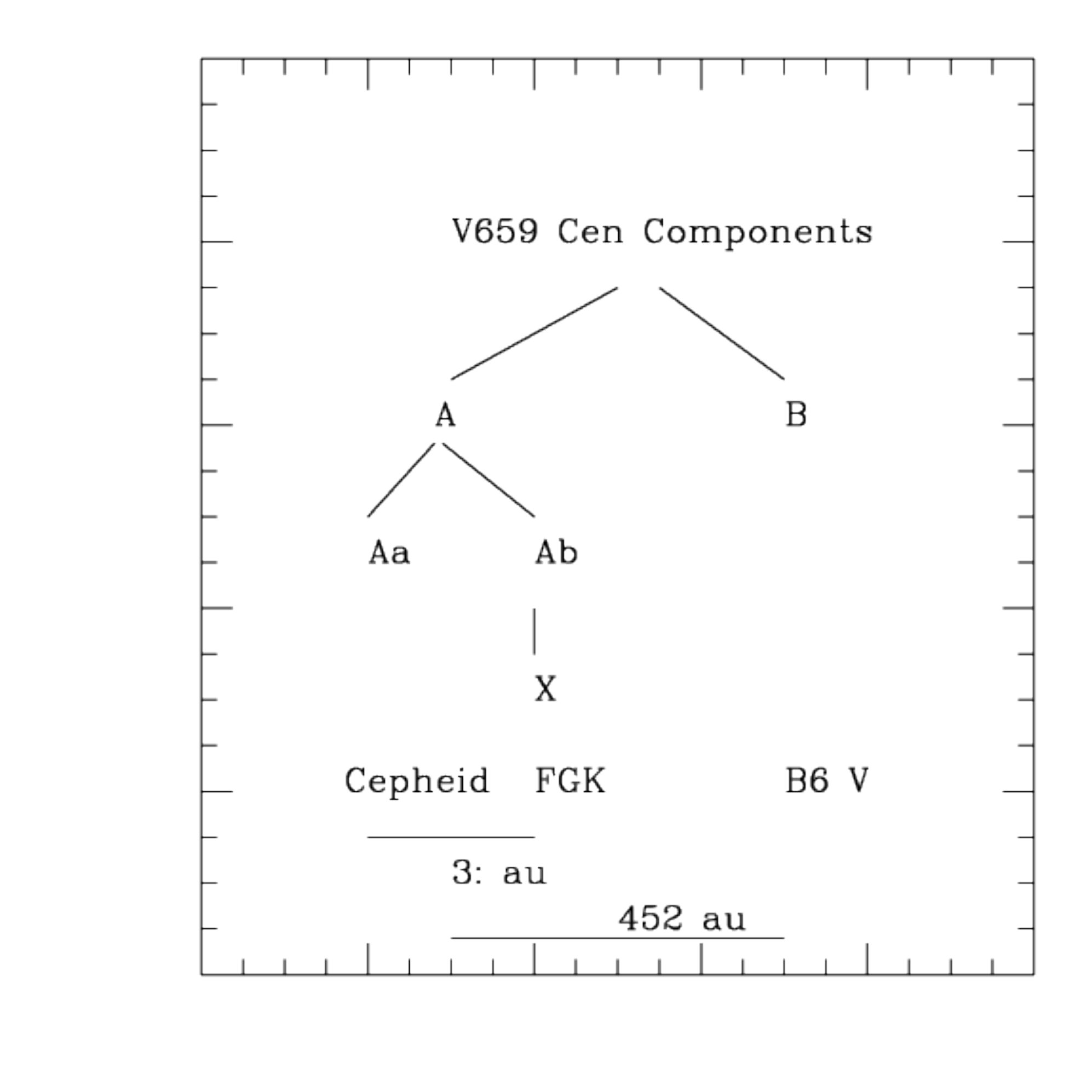}{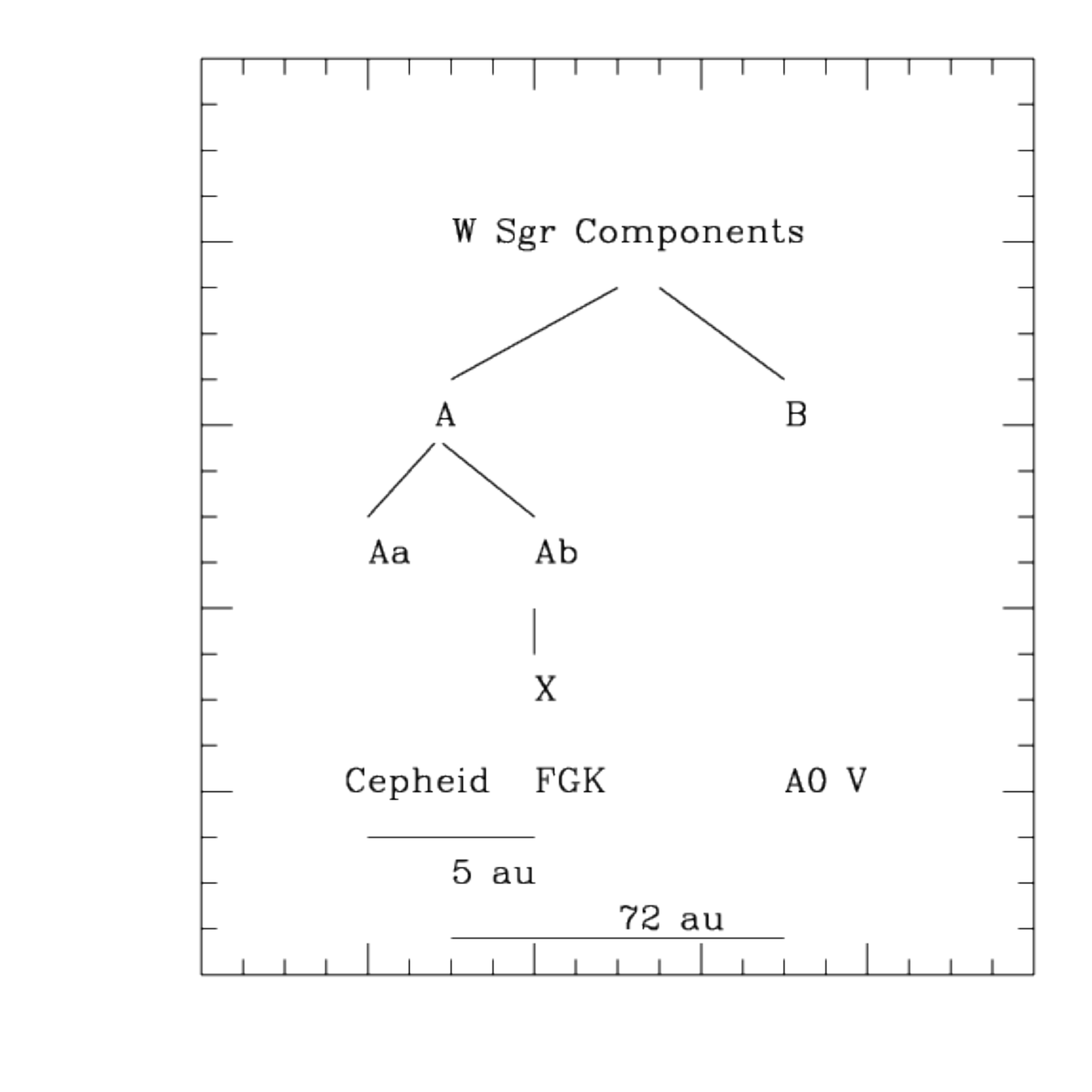} 
\caption{a. (left) The V659 Cen system.  The diagram of the components indicates the Cepheid and the 
spectral types of the close and distant binary.  Star Ab is a low-mass star (FGK) which produces the X-rays.  The
separations of the systems in au is indicated at the bottom.  (The separation in the Aa--Ab binary is only
an estimate.)  b. (right) The W Sgr system, with the same notation as a.   
\label{v659.wsgr}}
\end{figure}

{\bf V659 Cen}:  The Cepheid is in a multiple system and its components are only identified using 
a number of approaches.  A resolved companion at 0.6 $\arcsec$ 
(452 AU)  was found in a HST WFC3 survey 
of Cepheids (Evans et al. 2013).  The system  was found to be an X-ray source in the XMM-Newton 
survey of possible resolved companions (Evans, et al 2016b) with log L$_X$ = 29.51 ergs s$^{-1}$
at a pulsation phase of 0.14.
In the summary discussion of the HST WFC3 survey, Evans et al. (2020a),  found that 
systems with a resolved companion also have an inner spectroscopic binary.  Evidence for an 
inner binary in the V659 Cen system discussed there include orbital motion in velocities and 
possible orbital motion in Hipparcos proper motions. Can we identify the source of the X-rays in 
the triple system?  In comparison with Cepheids which have themselves been detected in X-rays 
(Table~\ref{x.c.obs}), V659 Cen has a much larger X-ray flux, particularly at the phase of the
observation.  V659 Cen B, the hottest star has a spectral type of B6 V from the ultraviolet 
spectrum (Evans et al 2020a).  The same study discusses an HST STIS ultraviolet 
spectrum oriented to resolve the Cepheid and the 0.6$\arcsec$ companion, which shows that the 
hottest star in the system   is the resolved companion V659 Cen B.  Could the X-rays be 
produced by that star?  X-rays are produced by O and early B stars (Berghoefer et al. 1997;
Naze, et al. 2011).  However the dividing line for X-ray producers is approximately B3 V.  
V659 Cen B is cooler than that and unlikely to produce X-rays.  The spectroscopic binary
companion V659 Cen Ab is a lower mass star, and hence should be able to produce X-rays, 
and, indeed the X-ray flux is reasonable for an F, G, or K star. 
The components are  summarized in Fig~\ref{v659.wsgr}a.
%Stis spectrum 

{\bf R Cru}  X-ray flux from the R Cru system was discovered by Evans, et al (2016b) in an 
XMM-Newton survey of possible resolved companions. The HST observations showed two possible 
companions as sources of the X-rays, one at 1.9$\arcsec$ and a closer spectroscopic binary.
A shallower Chandra exposure localized the X-rays to the spectroscopic binary 
companion (Evans, et al 2020a).  (Removal of the  1.9$\arcsec$ companion also removes the 
most discordant point in the CMD of companions in Fig. 6 in Evans, at al. [2020a]).
In addition, there is a star at 7.7$\arcsec$ (Evans, et al. 2020a; Kervella, et al. 2019b).
It was considered a likely  companion from Gaia DR2 data. However, in the EDR3 data, both 
the parallax and the proper motion do not match those of the Cepheid as closely and they are less
likely to be physically related.   
The system thus contains a Cepheid R Cru Aa, 
%a resolved companion R Cru B at a separation of 1580 AU,  
and the likely spectroscopic binary companion R Cru Ab. 
(Evans, et al 2020a). 
%summarized in Fig~\ref{v659.rcru}b.
% and a resolved companion at 7.7$\arcsec$, R Cru C.  
The X-ray luminosity  in Table~\ref{x.com.obs} is from the deeper XMM-Newton observation,

{\bf S Mus} X-rays were similarly discovered in the S Mus system in the XMM-Newton observation
by Evans, et al (2016b), which was followed by a Chandra observation to localize the X-rays 
(Evans, et al 2020a).  In this case, the X-rays come from a spectroscopic binary with a period
of 505 days made up of the Cepheid and a B3 V companion.  A companion this hot can produce 
X-rays through wind shocks, so the most likely interpretation
is that  the hot companion is responsible for the X-rays.  S Mus is thus the one Cepheid  
which is not an X-ray test for a low mass companion.  The X-ray luminosity in Table~\ref{x.com.obs}
is from the XMM-Newton observation.

{\bf W Sgr} X-ray flux was found at the location of the Cepheid by XMM-Newton, listed in the source 
catalog 4XMM-DR11 \footnote{http://xmm-catalog.irap.omp.eu/sources}.  
There are in fact 2 XMM-Newton observations of W Sgr.   In the second, the star is 
at the border of a chip, hence the flux is less precise.  The stacked observation has 
 a flux of  3.0 $\pm$ 0.4   x 10$^{-14}$   ergs s$^{-1}$ cm$^{-2}$ in 0.2-12 keV.
This is log L$_X$ = 29.78 at 409 pc.  

%W Sgr: companion  rosat /home/evans/xmm/cyc19/ceph/wsgr
 W Sgr is part of a  triple system (Evans, Massa, and Proffitt 2009).  The hottest companion to the
Cepheid, W Sgr B, has spectral type A0 V (Evans, et al, 2013) and was resolved from the spectroscopic 
binary in an HST STIS spectrum (Evans, Massa, and Proffitt 2009) at a projected distance of 72 AU.  
The spectroscopic binary Aa + Ab has separation of 5.0 AU (Benedict, et al. 2007). 
 Only an upper limit could 
be obtained for  the mass and spectral type of 
the companion from the STIS spectrum ($<$1.4 M M$_\odot$ 
and later than F5 V).
This is consistent with the X-ray luminosity.  
Components are summarized in Fig~\ref{v659.wsgr}b.

%%%%%other survey stars?

\subsection{X-rays from Main Sequence Stars}

Studies of activity in main sequence stars have been a very important area of X-ray study.  The 
X-ray luminosities in  the sequence of open clusters of different ages are summarized, for 
instance, in Prebisch and Feigelson (2005).  The age of a stars found in the instability strip  
 depends on their  mass and hence their pulsation period.
 A typical age is 50 Myr
as discussed by Bono, et al. (2005).  This is between the ages of the 
Orion Nebula Cluster and the Pleiades, with the $\alpha$ Per cluster being a good representation 
of stars of this age.  Its age is estimated to be 50 Myr (Meynet et al. 1993) to 90 Myr 
(Stauffer, et al. 1999).    
A study was made  by  Randich, et al. (1996)
of ROSAT observations of the full cluster found 
%In fact, the    results of Randich, et al. is 
%actually stronger:
 that at a depth of log L$_X$ to be 29.5 ergs s$^{-1}$, 76\% and 79\% of 
G and K stars were detected.  Even for F stars 75\% were detected although early F stars are 
not strong X-ray producers. 
A more recent study of a deep XMM-Newton observation of the $\alpha$ Per 
cluster is presented by Pillitteri, et. al. (2013).  They find a mean X-ray luminosity log L$_X$ 
to be 29.63 ergs s$^{-1}$ for F main sequence  stars, 29.74 ergs s$^{-1}$ for G dwarfs, and 
29.56 ergs s$^{-1}$ for K dwarfs.  M dwarfs are fainter, and are not expected to be detected 
in the present study.    A line is included at 29.6 ergs s$^{-1}$ on Fig~\ref{ul}
to indicate the
mean level of F, G, and K stars.  
 There is, of course, a range of X-ray luminosities for 
any mass or spectral type.  This is partly because of variation of rotation velocity between
stars.  In addition, cool stars have activity cycles.  However, even at a depth of 
log L$_X$  29.5, Randich, et al. find they detect three quarters of F, G, and K stars or better.

\section{ Low-Mass Companions of Cepheids} 

A sample of 20 Cepheids  observed in X-rays has been assembled  
from observations made for a variety of purposes. 
The X-ray luminosity level has been established for
quiescent phases from $\delta$ Cep and Polaris  at  
log L$_X$ $\simeq$ 28.7 ergs s$^{-1}$. On the other hand, deeper exposures for {\it l} Car and 
$\eta$ Aql have not detected the Cepheid at log L$_X$ 28.2 ergs s$^{-1}$.

 For the Cepheids W Sgr, V473 Lyr, V659 Cen, and R Cru  a young low mass
companion dominates the X-ray range. For S Mus the X-ray flux is most likely 
produced by an early B hot companion, which reduces the sample to 19 
Cepheids to look for low mass companions.  
 For the remaining 15 Cepheids, the upper limits
indicate that there is {\it not} a low mass companion.  
Thus, only 21\% of the Cepheids
clearly have a low-mass companion.  This fraction would be 
increased slightly if 2 stars with upper limits above or on the dividing line
were removed from the sample.
The sample has some limitations.  Systems with 
periods shorter than a year are not present in Cepheid samples because they would
 have disappeared due to RLOF, particularly at the tip of the red
giant branch.  The X-ray exposure depth was set to detect F, G, and K main sequence stars 
at the age of the Cepheids.  Thus M companions would not have been detected.  In 
X-ray studies of main sequence stars at this age, three quarters of F, G, and K 
companions would have been detected at this exposure depth.  This correction would
raise the fraction of systems with low-mass companions to  28$_{-9}^{+13}\%$
(errors from binomial statistics). This is clearly much 
lower than a random selection of companions from the IMF (Chabrier 2003; 
Moe and Di Stefano 2017),  at least for systems 
with separations greater than about 1 au.

%discuss binary properties of sample  

%no more low mass companions than xxx 30\% massive compnaions: paper 1 or crave

\section{Discussion}

\subsection{Binary/Multiple Fraction of Cepheids}
All methods of identifying Cepheid companions have some limitations.  The X-ray observations in 
the current paper do not detect companions of spectral types earlier than F or later than K.  
However they identify companions at any separation.  Similar properties are true for 
ultraviolet (UV) surveys: they identify companions earlier than mid-A at any separation (Evans, 
et al. 2013).  Radial velocities (Evans et al. 2015) and Gaia proper motions (Kervella, et al. 
2019a), on the other hand, are sensitive to a wider range of spectral types but detect
short period small separation systems, but not longer period systems.  Velocity studies are 
also much more sensitive for sharp-lined stars such as Cepheids than for broad-lined hot 
stars.  

Ultimately,  results for X-ray studies and ultraviolet studies of Cepheids
need to be combined into
the binary/multiple star fraction.  The X-ray fraction (28\%) 
 and the UV 
fraction (21\%; Evans 1992) are comprehensive for the companion spectral types they cover. 
However, Cepheids like other intermediate and massive stars are frequently found in systems
with more than two members, and thus results from these two ``detection wavelength'' approaches
would sometimes overlap.  We can make a rough estimate of this from this study, in that
of the 4 Cepheids with a late type companion two (V659 Cen and W Sgr) were already known
to be in multiple systems from UV studies.  That is, only half the  low-mass companions 
(14\%) are new systems in the total, resulting in 35\% of Cepheids in binary or multiple 
systems from the  combined  X-ray and UV studies.  This is,
of course, a lower limit since it does not represent all companion spectral types.  
We can further make rough estimates of the companions which are left out of these two spectral type 
regions.  The UV spectra identify all massive companions but have serious 
incompleteness starting at mid-A spectral types, corresponding to a mass of approximately 
1.9 M$_\odot$.   Table~\ref{mas.cov} summarizes information about these spectra type regions (bins).  
The top row lists the spectral types of the hot and cool boundaries of the  UV and
X-ray surveys. Corresponding masses are listed in the next line
 taken from Drilling and Landolt (2000).
Below the masses, the entries list information for the four bins.   The 
 mass ratio is shown for the mass range in the bin. The bottom line shows for 
percentage of companions measured in bins 1 and 3.   The mass ratios in line 3 are 
quite similar for the X-ray and UV regions and also for the regions not covered.  Following the results
from the IMF (Chabrier 2003), the region in bin 2
 not sampled in X-ray or UV is expected to 
have somewhat fewer stars than  the sampled regions and hence  fewer binary companions.  
Thus the missing region in bin 2 would probably
not double either the fraction in bins 1 and 3, or the combined fraction 
 (35\%), but it would add significantly.  Similarly, the missing M stars in bin 4 would 
substantially increase the fraction.  However, we have shown here that the fraction of 
cool companions in binary systems is less than predicted by a field IMF.
  In sum, the binary fractions in the bins 1, 2, and 3
are 21\%, $<$21\%, and 28\%.
A simple total is $<$70\%. It would be reduced somewhat by an overlap of UV and X-ray 
companions in multiple systems. 
However, it would be increased by   a substantial but unknown fraction of M stars   in bin 4.

\begin{deluxetable}{lccccccccc}
%\footnotesize
\tablecaption{Mass/Spectral Type Regions for Companion Detection  \label{mas.cov}}
\tablewidth{0pt}
\tablehead{
 \colhead{}  &  \colhead{Sp Ty UV}  &  \colhead{} & \colhead{Sp Ty UV}   
 & \colhead{}  & \colhead{Sp Ty X} &
 \colhead{}  &  \colhead{Sp Ty X}   & \colhead{}  & \colhead{}  \\
 \colhead{} &  \colhead{Hot}  & \colhead{}   &  \colhead{Cool}  & \colhead{} & \colhead{Hot} 
 & \colhead{}  &  \colhead{Cool}  & \colhead{}  & \colhead{}  \\
 \colhead{Bin} &  \colhead{}  & \colhead{1}   &  \colhead{}  & \colhead{2} & \colhead{} 
 & \colhead{3}  &  \colhead{}  & \colhead{4}  & \colhead{}  \\
}
\startdata
 Spectral Type  &  B5 & &    A5  & &   F5 & &  M0 & &   M5  \\
Mass M $_\odot$  &  5.9  & &  1.9  & &    1.4   & &   0.51  & &    0.21  \\
Mass Ratio & &    3.1   & &   1.4   & &   2.8  & &    2.4   &   \\ 
Companion \% & &   21   & &  --    & &   28    & &  --      &   \\ 
\enddata

\end{deluxetable}

The second binary/multiple detection technique is through orbital motion, either with radial 
velocities or with proper motions.  Evans et al. (2015) examine orbital motions for the 40 brightest
Cepheids N of -20$^o$ from two CORAVEL studies.  They find a binary fraction of 29\% with 
orbital periods between 1 and 20 years where the sample is most complete.  This rises  to
35\% for all periods greater than 1 year for the nearest 40 stars.  There is serious incompleteness 
for long orbital periods and low mass ratios.  Since from this X-ray study, companions F5 V or cooler 
are likely to be at least half the binary fraction, 
the incompleteness rises substantially.  
 Kervella, et al. (2019a) have compared Gaia DR2 
proper motions with those from Hipparcos to identify deviations resulting from orbital 
motion (proper motion anomalies).  Using their criterion for a detection of the ratio of
proper motion difference to signal to noise of 3, for the nearest 100 Cepheids, the binary 
fraction is 32\%.  This fraction doubles adding in binaries identified by other means, bringing it
close to the estimate of $~$70\% above. 

%   Msun: M$_\odot$ 

Wider companions in orbits which would not be identified either from velocities or proper motions 
 also exist.    The challenge here is that as larger separations 
from the Cepheid are searched, a field star is more likely to be included in the list of 
companion candidates.  This was tested with X-ray observations of a subset of possible resolved 
companions (Evans, et al. 2016b).  No  companion candidates with separations $>$5$"$ 
or 4000 au were confirmed to be  physical companions.  There are six systems with wider 
companions or possible companions in the HST survey (Evans, et al. 2020a), but most have a 
hot companion or are in spectroscopic binary systems, so they are already counted and
do not add to the list of binary
or multiple systems.  
Gaia DR2 and EDR3 parallaxes and proper 
motions have also
been used to investigate companions at wider separations (Kervella, et al. 2019b; Breuval 2021).  
This is a very promising approach, but so far for the  Cepheids within about 1 kpc, the 
   companion candidates all have sizable errors in EDR3 parallaxes.

The  most important feature in comparing the Cepheid binary fraction with that of B stars from which 
Cepheids evolved is RLOF for short period B binaries.  The effects are particularly striking in O stars
(Sana, et al. 2012).  Moe and Di Stefano (2017) find that only 75\% of mid-B stars will evolve into 
Cepheids.  This means our fraction of Cepheids with low-mass companions (28\%; Table~\ref{mas.cov}) 
corresponds to a fraction of 37\% of B stars. %This is very close to the companion fraction of 
%B stars in Tr 16 derived from a similar techique.  

 The fraction of low--mass companions of Cepheids  can be compared with that of ``late B"
stars in Tr 16 (39\%) from a similar X-ray technique (Evans, et al. 2011).  This is very 
close to the Cepheid fraction (37\%).
The fraction in Tr 16 might  be somewhat higher since 
the cluster is younger than the Cepheids, and 
hence low--mass stars are more X-ray active and more easily detected.  Furthermore,
the Cepheid sample is limited to binaries with periods longer than a year.    However the 
similarity of the fractions indicates that in both the occurrence of low mass companions
is lower than would be predicted by random sampling from a field IMF.

%XXXX  Low mass cepheids fill in range in fig 1 of moe and distefano

%sf processes in binary different from field

\subsection{Implications for Star Formation}  

\begin{figure}
\plotone{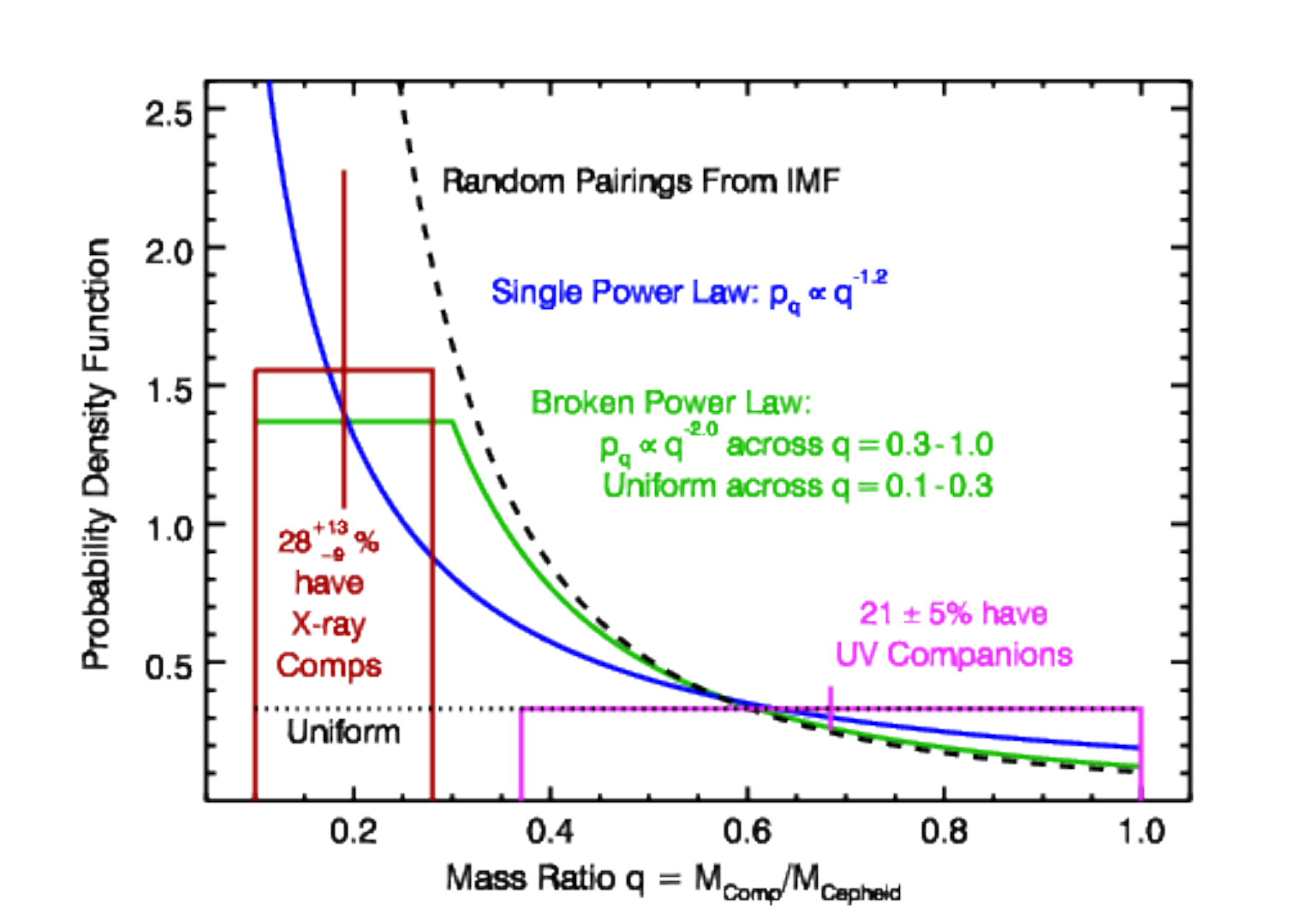} 
\caption{ Our X-ray survey demonstrates that 28$_{-9}^{+13}\%$ of Cepheids have unresolved 
late-F/G/K companions across $q$ = 0.10 – 0.28 (red). Meanwhile, an earlier UV survey 
(Evans 1992) found that 21\% $\pm$ 5\% of Cepheids have unresolved B/early-A companions 
across $q$ = 0.37 – 1.0 (magenta). The measured mass-ratio distribution is inconsistent 
with both a uniform distribution (dotted) and random pairings of the IMF (dashed). We fit 
a power-law distribution with an intermediate slope of $\gamma$ = $-$1.2 $\pm$ 0.4 across 
$q$ = 0.1 – 1.0 (blue). The data are also consistent with the segmented power-law model 
adopted in Moe \& Di Stefano (2017; green). In total, we find that 57\% $\pm$ 12\% of 
Cepheids have companions across $q$ = 0.1 – 1.0 and $a$ = 1 - 1,000 au.
\label{st.form}}
\end{figure}

Unresolved companions to Cepheids provide a unique probe into the properties of intermediate 
mass binaries across intermediate separations, which helps to constrain binary formation  
models. The unresolved companions must be wider than a $>$ 1 au to avoid RLOF 
with the Cepheid supergiant primaries, and must also reside within a $<$ 1000 au; otherwise we 
would have resolved the companions in our previous HST imaging campaign (Evans et al. 2020a). 
Our detected X-ray companions span late-F/G/K dwarfs, corresponding to masses 
M$_{comp}$ = 0.5-1.4 M$_\odot$. For a typical Cepheid primary mass of M$_{Cepheid}$ $\approx$ 5 M$_\odot$, 
the binaries correspond to mass ratios q = M$_{comp}$ /M$_{Cepheid}$ = 0.10 - 0.28. After correcting 
for incompleteness of  FGK stars that do not produce a detectable X-ray flux as described 
above, we conclude that 28$_{-9}^{+13}\%$ of Cepheids have companions across a = 1 - 1000 au 
and q = 0.10 - 0.28 (red data point in Fig.~\ref{st.form}), where the uncertainties derive from binomial 
statistics. For a larger sample of 76 Cepheids, 16 exhibited a UV excess from unresolved 
B/early-A dwarfs spanning M$_{comp}$ = 1.9 – 5.9 Msun (Evans 1992). We thus find that 21\% $\pm$ 5\% 
of Cepheids have companions across a = 1 - 1000 au and q = 0.37 – 1.00 (magenta data point
 in Fig.~\ref{st.form}). We are incomplete to late-A/early-F dwarf companions, which span the narrow 
mass-ratio interval q = 0.28 – 0.37.

As shown in Fig.~\ref{st.form}, companions to Cepheids are skewed toward smaller mass ratios. Given the 
measured occurrence rate of B/late-A companions to Cepheids via the UV excess method and 
assuming a uniform mass-ratio distribution (dotted line in Fig.~\ref{st.form}), we would have expected 
only 6\% of Cepheids to have late-F/G/K companions across q = 0.10 – 0.28. This prediction 
is discrepant with our empirical measurement of 28$_{-9}^{+13}\%$ at the 
2.7$\sigma $ level. Conversely, if we instead assumed that binaries were drawn from random 
pairings of the IMF (dashed curve in Fig.~\ref{st.form}), we would have expected 87\% of Cepheids to 
have companions across q = 0.10 – 0.28, which is even more inconsistent with our measurement 
at the 4.5$\sigma$ level. The true mass-ratio distribution is between these two slopes. 
By fitting a single power-law distribution $p_q$ $\propto$ $q^{\gamma}$, we measure 
$\gamma$ = $-$1.2 $\pm$ 0.4 across $q$ = 0.1 – 1.0 (blue curve in Fig.~\ref{st.form}).

Close companions (a $<$ 1 au) to mid-B stars follow a uniform mass-ratio distribution, 
indicating they coevolved via shared accretion in a circumbinary disk, while wide companions 
(a $>$ 1000 au) are weighted toward extremely small mass ratios, nearly consistent with 
random pairings drawn from the IMF, suggesting the components fragmented and subsequently 
accreted fairly independently (Abt et al. 1990; Kouwenhoven, et al. 2007; Kobulnicky and 
Fryer 2007; Moe and Di Stefano 2017). Across intermediate separations, both long-baseline 
interferometry (Rizzuto, et al. 2013) and decomposition of binaries from high-resolution 
spectra (Gullikson, et al. 2016) demonstrated that the mass-ratio distribution is skewed 
toward small mass ratios. However, these techniques are insensitive to companions below 
q $<$ 0.3. Our survey yields the first robust census of low-mass companions to intermediate-mass 
stars across intermediate separations, confirming earlier indications that the mass-ratio 
distribution is skewed toward small mass ratios but nonetheless still top-heavy compared 
to random pairings drawn from the IMF. Thus both disk and core fragmentation and accretion 
lead to a mixed population of intermediate-period binaries.

Moe \& Di Stefano (2017) adopted a segmented power-law mass-ratio distribution with the 
parameter $\gamma_{\rm largeq}$  describing the slope across large mass ratios $q$ = 0.3 – 1.0 
and $\gamma_{\rm smallq}$ across $q$ = 0.1 – 0.3. They fitted both power-law slopes as a 
function of primary mass and orbital separation based on a combination of datasets, 
interpolating over the gaps in the observations. For intermediate-period companions to 
5M$_\odot$ primaries, they fitted $\gamma_{\rm largeq}$ =$-$2.0 and $\gamma_{\rm smallq}$ = 0.0 
(green curve in Fig.~\ref{st.form}). This distribution is significantly skewed toward small mass 
ratios, nearly consistent with random pairings of the IMF across $q$ = 0.3 – 1.0, but 
then flattens to a uniform distribution below $q$ $<$ 0.3. Their broken power-law model 
is also consistent with our measurements.

We can now determine the overall unresolved binary fraction of Cepheids. By interpolating 
our best-fit power-law model across the mass ratio gap where both X-ray and UV methods are 
insensitive, we expect an additional 8\% of Cepheids to have late-A/early-F companions 
across $q$ = 0.28 – 0.37. Thus 57\% $\pm$ 12\% of Cepheids have companions across 
$q$ = 0.1 – 1.0 and $a$ = 1 - 1000 au. This is consistent with expectations from mid-B 
binaries. Moe \& Di Stefano (2017) estimated that 85\% of 5$_\odot$ MS primaries have 
companions above $q$ $>$ 0.1, of which 70\% have intermediate separations spanning 
$a$ = 1 – 1,000 au. Hence, 0.85$\times$0.70 = 60\% of mid-B stars have companions 
across intermediate separations, nearly identical to our Cepheid result.

\begin{figure}
\plotone{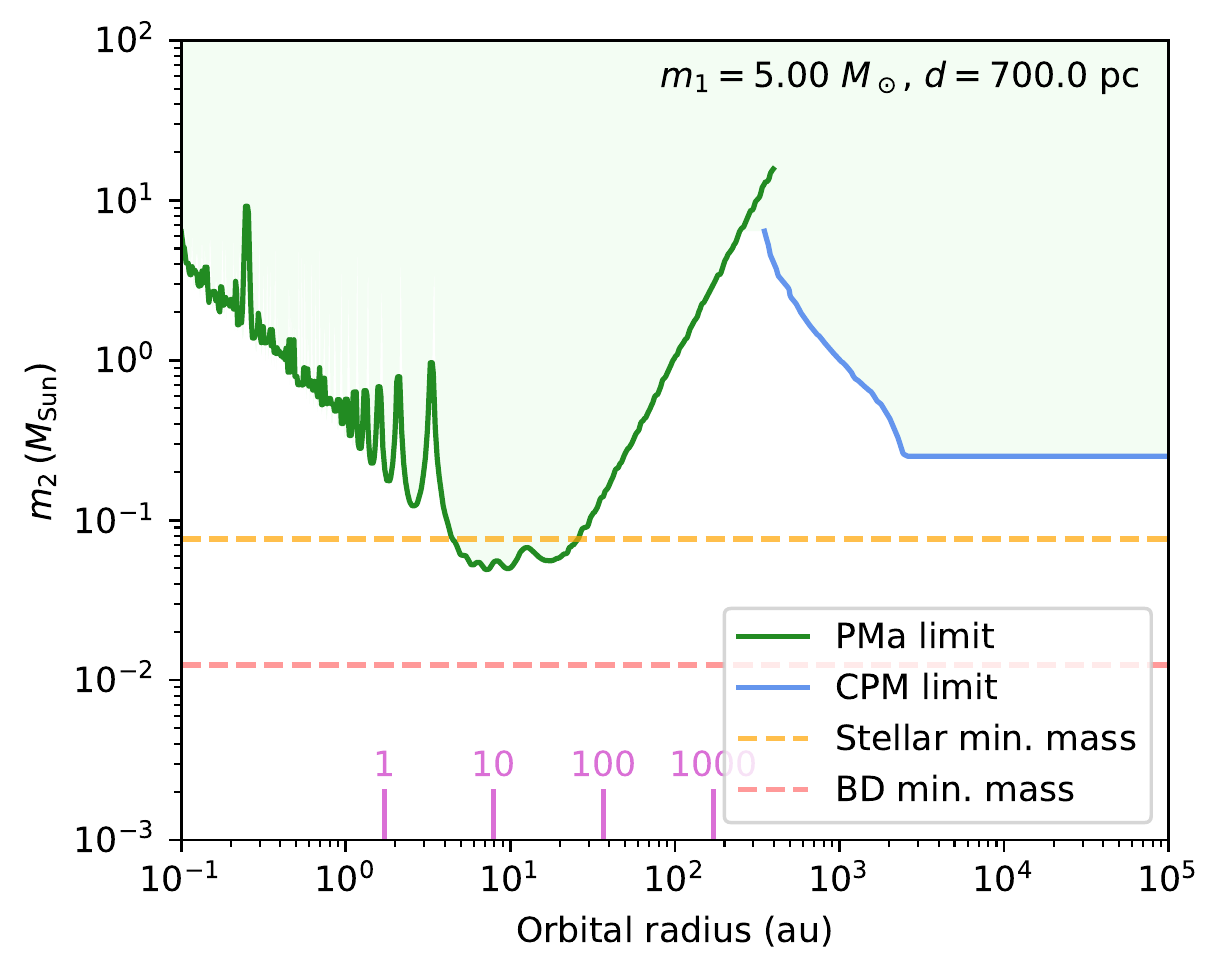} 
\caption{Companion Detection Sensitivity for Cepheids from Gaia EDR3.  Sensitivity for 
companion mass is shown as a function of orbital radius and orbital period.  Green line: limit
from proper motion anomaly; blue line: limit for common proper motions; yellow dashed line:
stellar mass limit; pink dashed line: brown dwarf limit; orbital period in years 
is shown by the pink lines on the x-axis.  Cepheid parameters are 5.0 M$_\odot$ 
and 700 pc,  typical parameters in this study.   The green region above the stellar limit
indicates that stellar companions at the low-mass stellar limit will be detected for orbital
periods between about 3  and 100 years.   
\label{gaia.sens}}
\end{figure}

 The X-ray studies here demonstrate that the fraction of F, G, and K companions is smaller 
than would be produced by random pairings of the IMF (Fig.~\ref{st.form}).  To complete
the understanding of companion distribution, we need to know the form of the distribution 
of M star companions.  A recent study of b Cen, a 6-10 M$_\odot$ binary demonstrates 
that even a planet can exist around a massive star (Janson, et al. 2022), providing the 
need to hunt for even smaller objects.  Observing a sample of M stars to explore the mass
distribution would require very long exposures in X-rays.  
  Kervella et al (2022) discuss the use of Gaia EDR3 proper motions
and parallaxes to identify companions 
around stars within 100 pc through proper motion anomalies (PMa; orbital motion) and common proper
motion pairs (CPM).  This sample includes relatively few massive stars, however 
they find that 45\% of the sample
have an indication of binarity and 7\% have bound CPM candidates. 
They estimate that as many as 70\% of Cepheid-mass stars could have PMa and CPM companions.
Fig~\ref{gaia.sens} shows the detection limits for a typical Cepheid in this study (5M$_\odot$ and
700 pc) in EDR3.  Stars  down to the low-mass limit should be detected for orbital periods
between about 3 and 100 years.  This limit will improve substantially in future Gaia releases, 
to approximately 4 $\times$ better in DR4 ($~$2024) and 13    $\times$ better in DR5 ($~$2027).

\section{Summary}

In this section  we summarize the main points of this study.

\subsection{l Car and Polaris} As shown in Fig~\ref{ul},  {\it l} Car  has an 
X-ray luminosity which is below  the quiescent phases of $\delta$ Cep, and certainly
below the X-ray burst at maximum radius in   $\delta$ Cep.  Because {\it l} Car has a
longer period than  $\delta$ Cep, we know it has a higher luminosity and mass, as 
well as reaching cooler minimum temperatures. Any of these could affect X-ray 
production, though it is not obvious that they would affect the convective surface to
disrupt magnetic activity.  While the observation was carefully timed for maximum radius,
in a 35$^d$ Cepheid, the 80 ksec exposure covers only 
3\% of the phase, so a phase-restricted 
burst could have been missed.  

For Polaris, the X-ray luminosity is comparable to that of  $\delta$ Cep in its 
quiescent phases.  There is no indication of an X-ray burst even though much of 
the pulsation cycle has been covered.  The observation discussed here alone 
covers a phase range of 0.48 to 0.69.  
Polaris pulsates in the first overtone mode and has a very 
low pulsation amplitude, which could  alter the X-ray--pulsation relation
as compared with $\delta$ Cep (full amplitude and fundamental mode).

\subsection{The Pulsation Cycle}  A major motivation for the series of X-ray observations of
Cepheids follows from the X-ray burst in $\delta$ Cep.   Fig~\ref{ul} shows that neither 
$\eta$ Aql nor {\it l} Car share this behavior.  As discussed above  {\it l} Car differs 
from  $\delta$ Cep in physical properties which might account for this.    $\eta$ Aql, on the
other hand, has a period very similar to  $\delta$ Cep, and hence,  similar luminosity, 
mass, and temperature cycle.  At this point, questions remain about the X-ray behavior of 
both stars.  Polaris has no indication of a phase-related X-ray increase.

\subsection{Low--Mass Companions}  
Chandra and XMM-Newton observations of a sample of 20 
Cepheids finds that 28\% have a low mass companion.  
The fraction of Cepheids with low-mass companions is very similar to that predicted 
from mid-B stars.
This sample identifying F, G, and K spectral type companions can be combined 
with a previous survey in the 
UV which identifies B and early A companions. 
% The result, including a correction for spectral 
%types between these  is that the fraction of Cepheids in binary or 
%multiple systems can approch 70\%.   
Using a Moe and Di Stefano segmented power law to fit the data, 
 57\% $\pm$12\% have companions with the mass ratio q $>$ 0.1 the separation a $>$ 1 au.  
This is the first survey of intermediate mass stars that reaches to mass ratios this small.
The mass ratio distribution falls between a  uniform distribution  and random pairings from IMF, 
that is between formation from 
 shared accretion in a  circumbinary disk and fragmentation and independently accreted
components.  

%Comparison with a sample of progenitors 
%(``late B'' stars in Tr 16) shows that this fraction has only been mildly altered by 
%evolution.  Thus the massive low--mass ratio is   distinctly different in binary systems and 
%in the field,   

\section{Acknowledgments}

This research is 
based on observations obtained with XMM--Newton, an 
ESA science mission with instruments and contributions directly 
funded by ESA Member States and the USA (NASA).

Support was provided to NRE by the Chandra X-ray Center NASA Contract NAS8-03060.
The observations were associated with program 84051
with support for this work from  
NASA Grant 80NSSC20K0794. JJD was supported by NASA contract NAS8-03060 to the 
 Chandra X-ray Center and thanks the Director, Pat Slane, for continuing advice and support.
HMG was supported through grant HST-GO-15861.005-A from the STScI under NASA contract NAS5-26555.
P.K. and L.B. acknowledge funding from the European Research Council (ERC) under the
 European Union's Horizon 2020 research and innovation program (projects CepBin, grant 
agreement No 695099, and UniverScale, grant agreement No 951549).
This work has made use of data from the European Space Agency (ESA) mission {\it Gaia} 
(\url{http://www.cosmos.esa.int/gaia}), processed by the {\it Gaia} Data Processing and 
Analysis Consortium (DPAC, \url{http://www.cosmos.esa.int/web/gaia/dpac/consortium}).
Funding for the DPAC has been provided by national institutions, in particular the 
institutions participating in the {\it Gaia} Multilateral Agreement.

The SIMBAD database, and NASA’s Astrophysics Data System Bibliographic Services
were used in the preparation of this paper.

%\clearpage


\begin{thebibliography}{}

\bibitem[\protect\citeauthoryear{abt etal}{1990}]{abt90}
Abt, Helmut A.,  Gomez, Ana E., Levy, Saul G. 1990, ApJS, 74, 551

\bibitem[\protect\citeauthoryear{anderson etal}{2015}]{and15}
Anderson, R. I., Sahlmann, J., Holl, B., et al. 2015, ApJ, 804, 144

\bibitem[\protect\citeauthoryear{anderson etal}{2016a}]{and16a}
Anderson, R. I., M\'erand, A., Kervella, P., et al. 2016a, MNRAS, 455, 4231 

\bibitem[\protect\citeauthoryear{anderson}{2016b}]{and16b}
Anderson, R. I. 2016b, MNRAS, 463, 1707

\bibitem[\protect\citeauthoryear{anderson}{2014}]{and14}
Anderson, R. I. 2014, A\&A, 566, L10

\bibitem[\protect\citeauthoryear{anderson}{2019}]{and19}
Anderson, R. I. 2019, A\&A, 623, A146

\bibitem[\protect\citeauthoryear{ayres}{2011}]{ayr11}
Ayres, T. 2011, ApJ, 738, 120

\bibitem[\protect\citeauthoryear{banyard etal}{2021}]{ban21}
Banyard, G., Sana, H., Mahy, L. et al. 2021, A\&A, accepted  


\bibitem[\protect\citeauthoryear{benedict etal}{2007}]{ben07}
Benedict, G. F., McArthur, B. E., Feast, M., et al. 2007, AJ, 133, 1810

\bibitem[\protect\citeauthoryear{berdnikov etal}{2000}]{ber00}
Berdnikov, L. N., Dambis, A. K., and Vozyakova, O. V. 
 2000, A\&AS, 143, 211


\bibitem[\protect\citeauthoryear{berghoefer et al}{1997}]{ber97}
Berghoefer, T. W., Schmitt, J. H. M. M., Danner, R., and Cassinelli, J. P. 1997, A\&A, 322, 167

\bibitem[\protect\citeauthoryear{bono et al}{2005}]{bon05}
Bono, G., Marconi, M., Cassisi, S., et al. 2005, ApJ, 621,966  

\bibitem[\protect\citeauthoryear{breuval}{2021}]{bre21}
Breuval, L  2021 PhD Thesis, Universit\'e Paris 

\bibitem[\protect\citeauthoryear{chabrier}{2003}]{cha03}
Chabrier, G. 2003, PASP, 115, 763

\bibitem[\protect\citeauthoryear{drilling landolt}{2000}]{dri00}
Drilling, J. S. and Landolt, A. U. 2000 in Astrophysical Quantities, ed. A. N. Cox (New York: Springer),
381

\bibitem[\protect\citeauthoryear{engle etal}{2017}]{eng17}
Engle, S. G., Guinan, E. F., Harper, G. M., et al. 2017, ApJ, 838, 67  (secret lives)

\bibitem[\protect\citeauthoryear{engle}{2015}]{eng15}
Engle, S. G. 2015, PhD Thesis, James cook University

\bibitem[\protect\citeauthoryear{evans}{1992}]{eva92}
Evans, N. R. 1992, ApJ, 384, 220 

\bibitem[\protect\citeauthoryear{evans etal}{2009}]{eva09}
Evans, N. R., Massa, D, and Proffitt, C. 2009, AJ, 137, 3700

\bibitem[\protect\citeauthoryear{evans etal}{2011}]{eva11}
Evans, N. R., DeGoia-Eastwood, K., Gagn\'e, M. et al. 2011, ApJS, 194, 13

\bibitem[\protect\citeauthoryear{evans etal}{2013}]{eva13}
Evans, N. R., Bond, H. E., Schaefer, G. H., Mason, B. D., Karovska, M., and 
Tingle, E.  2013, AJ, 146, 93 

\bibitem[\protect\citeauthoryear{evans etal}{2015}]{eva15}
Evans, N. R., Berdnikov, L., Lauer, J. et al. 2015 AJ, 150, 13 %  CRaV

\bibitem[\protect\citeauthoryear{evans etal}{2016}]{eva16}
Evans, N. R., Pillitteri, I., Wolk, S., et al. 2016b, AJ, 151, 108 % Pap IV

\bibitem[\protect\citeauthoryear{evans etal}{2016}]{eva16}
Evans, N., R., Bond, H. E., Schaefer, G. H., et al. 2016a, AJ, 151, 129 % pap ii

\bibitem[\protect\citeauthoryear{evans etal}{2018}]{eva18}
Evans, N. R. Karovska, M., Bond, H. E. et al. 2018, ApJ, 863, 187  %pol mass

\bibitem[\protect\citeauthoryear{evans etal}{2020b}]{eva20}
Evans, et al. 2020, AJ, 159, 121  % V473 Lyr

\bibitem[\protect\citeauthoryear{evans etal}{2020}]{eva20}
Evans, N. R., Guenther, H. M., Bond, H. E., et al. 2020a,
ApJ, 905, 81 %final

\bibitem[\protect\citeauthoryear{evans etal}{2021}]{eva21}
Evans, N. R., Pillitteri, I., Kervella, P. et al. 2021, AJ, 162, 92  %eta aql

\bibitem[\protect\citeauthoryear{fernie etal}{1995}]{fer95}
Fernie, J. D., Evans, N. R., Beattie, B., and Seager, S. 1995,
IBVS, 4148, 1

\bibitem[\protect\citeauthoryear{gallenne etal}{2016}]{gal16}
 Gallenne, A., M'erand, A., Kervella, P. et al. 2016, MNRAS, 461, 1451

\bibitem[\protect\citeauthoryear{gallenne etal}{2022}]{gal21}
 Gallenne, A., M'erand, A., Kervella, P. et al. 2021, A\&A, 651, A113

\bibitem[\protect\citeauthoryear{gullikson etal}{2016}]{gul16}
Gullikson, K.,  Kraus, A., and  Dodson-Robinson, S.  2016, AJ, 152, 40 

\bibitem[\protect\citeauthoryear{hocde etal}{2020a}]{hoc20a}
Hocd\'e, V, et al. 2020a, A\&A, 633, A47

\bibitem[\protect\citeauthoryear{hocde etal}{2020b}]{hoc20b}
Hocd\'e, V, et al. 2020b, A\&A, 641, A74

\bibitem[\protect\citeauthoryear{hocde etal}{2021}]{hoc21}
Hocd\'e, V, et al. 2021, A\&A, 651, A92

\bibitem[\protect\citeauthoryear{janson etal}{2022}]{jan22}
Janson, M, Gratton, R. Roder, L. et al. 2022  arXiv:2112.04833


\bibitem[\protect\citeauthoryear{kervella etal}{2022}]{ker22}
Kervella, P., Arenou, F., and Th\'evenin, F. 2022, arXiv:2109.10902v2

\bibitem[\protect\citeauthoryear{kervella etal}{2019a}]{ke19a}
Kervella, P. Gallenne, A., Evans, N. R., et al 2019a, A\&A, 623, A116 %I pn anom

\bibitem[\protect\citeauthoryear{kervella etal}{2019b}]{ke19b}
Kervella, P., Gallenne, A., Evans, N. R. et al. 2019b, A\&A, 623, A117 
%II: comoving wide comp

\bibitem[\protect\citeauthoryear{kobulnicky fryer}{2007}]{kob07}
Kobulnicky, H. A. and Fryer, C. L.   2007, ApJ, 670, 747

\bibitem[\protect\citeauthoryear{kouwenhoven etal}{2007}]{kou07}
 Kouwenhoven, M. B. N., Brown, A. G. A., Portegies Zwart, S. F., and Kaper, L.    
2007, A\&A, 474, 77



\bibitem[\protect\citeauthoryear{merand etal}{2006}]{mer06}
Merand, A.et al. 2006, A\&Ap, 453, 155

\bibitem[\protect\citeauthoryear{meynet etal}{1993}]{mey93}
Meynet, G., Mermilliod, J.-C., and Maeder, A. 1993, A\&AS, 98, 477

\bibitem[\protect\citeauthoryear{moe di stefano}{2017}]{moe17}
Moe, M. and Di Stefano, R. 2017, ApJS, 230, 15 

\bibitem[\protect\citeauthoryear{naze etal}{2011}]{naz11}
Naze, Y., Broos, P., Oskinova, L. et al. 2011, ApJS, 215, 10 

\bibitem[\protect\citeauthoryear{neilson etal}{2016}]{nei16}
Neilson, H. R., Engle, S. G., Guinan, E. F., Bisol, A. C., and Butterworth, N. 2016, ApJ, 824, 1

\bibitem[\protect\citeauthoryear{pillitteri etal}{2013}]{pil13}
Pillitteri, I., Evans, N. R., Wolk, S., and Syal, M. B. 2013 AJ, 145, 143  alp per

\bibitem[\protect\citeauthoryear{preibisch feigelson}{2005}]{pre05}
Prebisch, T. and Feigelson, E. D. 2005, ApJS, 160, 390

\bibitem[\protect\citeauthoryear{randich et al}{1996}]{ran96}
Randich, S., Schmitt, J. H. M. M., Prosser, C. F., and Stauffer, J. R. 1996, 
A\&A, 305, 785

\bibitem[\protect\citeauthoryear{rizzuto et al}{2013}]{riz13}
Rizzuto, A. C.,  Ireland, M. J., Robertson, J. G., et al.    2013, MNRAS, 436, 1694

\bibitem[\protect\citeauthoryear{sana et al}{2012}]{san12}
Sana, H., de Mink, S. E., de Koter, A. et al. 2012, Sci., 337, 444

\bibitem[\protect\citeauthoryear{seward}{2000}]{sew00}
Seward, F. D. 2000, in Astrophysical Quantities, ed. A. N. Cox (New York: Springer),
381

\bibitem[\protect\citeauthoryear{stauffer barrado}{1999}]{sta99}
Stauffer, J. R., Barrado y Navascues, D. et al. 1999, ApJ, 527, 219

\bibitem[\protect\citeauthoryear{szabados}{1989}]{sza89}
 Szabados, L. 1989, Mitt. Sternwarte Ungar Akad Wissen., 94, 1


\bibitem[\protect\citeauthoryear{szabados}{1991}]{sza91}
 Szabados, L. 1991, Mitt. Sternwarte Ungar Akad Wissen., 96, 123

\bibitem[\protect\citeauthoryear{taylor etal}{1997}]{tay97}
Taylor, M. M., Albrow, M. D., Booth, A. J., and Cottrell, P. L. 1997, MNras, 292, 662


 \bibitem[\protect\citeauthoryear{usenko}{2013}]{use13}
Usenko, I. A., Kniazev, A. Yu., Berdnikov, L. N., Kravtsov, V. V.,  and
Fokin, A. B. 2013,  AstL, 39, 432

\bibitem[\protect\citeauthoryear{usenko}{2014}]{use14}
Usenko, I. A., Kniazev, A. Yu., Berdnikov, L. N., 
Fokin, A. B.  and Kravtsov, V. V.  2014,  AstL, 40, 435

\bibitem[\protect\citeauthoryear{usenko etal}{2016}]{use16}
Usenko, I. A.,  Kovtyukh, V. V.,  Miroshnichenko, A. S., and Danford, S. 
2016, OAP, 29, 100
  



\end{thebibliography}
\end{document}